\documentclass[prd,nofootinbib,preprintnumbers,twocolumn,showpacs,floatfix,superscriptaddress]{revtex4-1}

\usepackage{amsmath}
\usepackage{amssymb}
\usepackage{graphicx}
\usepackage{color}
\usepackage{hyperref}

\newcommand{\bitem}{\begin{itemize}}
\newcommand{\eitem}{\end{itemize}}
\newcommand{\bwt}{\begin{widetext}}
\newcommand{\ewt}{\end{widetext}}
\newcommand{\beq}{\begin{equation}}
\newcommand{\eeq}{\end{equation}}

\newcommand{\bdm}{\begin{displaymath}}
\newcommand{\edm}{\end{displaymath}}
\newcommand{\bea}{\begin{eqnarray}}
\newcommand{\eea}{\end{eqnarray}}

\newcommand{\nn}{\nonumber}

\def\eq#1{{Eq.~(\ref{#1})}}
\def\eqs#1#2{{Eqs.~(\ref{#1})--(\ref{#2})}}

\def\fig#1{{Fig.~\ref{#1}}}

\def\Table#1{{Table~\ref{#1}}}

\def\sect#1{{Sect.~\ref{#1}}}

\def\app#1{{Appendix~\ref{#1}}}

\def\vev#1{\left\langle #1 \right\rangle}

\def\abs#1{\left| #1\right|}

\def\Tr{\!\mathop{\rm Tr}}

\newcommand\GeV{\text{GeV}}

\begin{document}

\title{Neutrino-axion-dilaton interconnection}
\author{Stefano Bertolini}
\email{stefano.bertolini@sissa.it}
\affiliation{INFN, Sezione di Trieste, SISSA,
Via Bonomea 265, 34136 Trieste, Italy\\[1ex]}

\author{Luca Di Luzio}
\email{luca.di.luzio@ge.infn.it}
\affiliation{Dipartimento di Fisica, Universit\`a di Genova and INFN, Sezione di Genova,
Via Dodecaneso 33, 16159 Genova, Italy\\[1ex]}

\author{Helena Kole\v{s}ov\'a}
\email{helena.kolesova@fjfi.cvut.cz}
\affiliation{Faculty of Nuclear Sciences and Physical Engineering,
Czech Technical University in Prague,
B\v{r}ehov\'a 7, 115 19 Praha 1, Czech Republic\\[1ex]}
\affiliation{Institute of Particle and Nuclear Physics,
Faculty of Mathematics and Physics,
Charles University in Prague, V Hole\v{s}ovi\v{c}k\'ach 2,
180 00 Praha 8, Czech Republic\\[1ex]}

\author{Michal Malinsk\'y}
\email{malinsky@ipnp.troja.mff.cuni.cz}
\affiliation{Institute of Particle and Nuclear Physics,
Faculty of Mathematics and Physics,
Charles University in Prague, V Hole\v{s}ovi\v{c}k\'ach 2,
180 00 Praha 8, Czech Republic\\[1ex]}

\author{Juan Carlos Vasquez}
\email{jcvasque@sissa.it}
\affiliation{SISSA/ISAS,
Via Bonomea 265, 34136 Trieste, Italy\\[1ex]}
\affiliation{ICTP,
Strada Costiera 11, 34014  Trieste, Italy\\[1ex]}
\affiliation{INFN, Sezione di Trieste, SISSA,
Via Bonomea 265, 34136 Trieste, Italy\\[1ex]}


\begin{abstract}

\noindent
We show that a recently proposed framework that provides a simple connection between Majorana neutrinos and 
an invisible axion in minimal scalar extensions of the standard electroweak model can be  naturally embedded in a classically scale-invariant setup. 
The explicit breaking of the scale invariance \`a la Coleman-Weinberg generates the Peccei-Quinn and electroweak scales. 
The spontaneous breaking of the chiral $U(1)_{PQ}$ triggers the generation of neutrino masses via Type-II seesaw and, at the same time, 
provides a dynamical solution to the strong CP problem as well as the axion as a dark matter candidate.
The electroweak and neutrino mass scales are obtained via a technically natural ultraweak limit of the singlet scalar interactions.  
Accordingly, a realistic and perturbatively stable scalar spectrum, possibly in the reach of the LHC, is naturally obtained. 
A very light pseudodilaton characterizes such a setting. The vacuum stability of the extended setup is discussed.

\end{abstract}

\maketitle

\section{Introduction}

In spite of the fundamental importance for the understanding of the electroweak symmetry breaking, the discovery of what appears to be the long-sought Higgs boson still leaves many issues of the Standard Model (SM) of particle interactions unaddressed. Laboratory and astrophysical observations give us an extremely detailed picture of massive neutrino and lepton mixing which clearly indicate the need for physics beyond the SM. Dark matter is required to account for more than 20\% of the mass of {the Universe}, where antimatter is a very rare component. From a theoretical point of view, the absence of new particles at the TeV scale raises the issue of the stability of the electroweak scale in the presence of hypothetical new heavy states associated to, e.g., grand unification or other high-scale dynamics (if nothing else gravity does).

In a recent paper~\cite{Bertolini:2014aia} we attempted to show that most of these issues can be organically addressed in a minimal renormalizable framework that just extends the Higgs sector of the SM.  While providing a structural connection among different open questions, the model naturally offers a stable spectrum of exotic scalar states at the TeV scale, thus opening the possibility of a test at the LHC.

The proposal in~\cite{Bertolini:2014aia} extends the model discussed in~\cite{Bertolini:1990vz,Arason:1990sg} which provided a connection between an invisible axion \`a la Dine-Fischler-Srednicki-Zhitnitsky (DFSZ)~\cite{Dine:1981rt,Zhitnitsky:1980tq} {and} the one-loop generation of neutrino masses \`a la Zee~\cite{Zee:1980ai,Wolfenstein:1980sy}. As explicit and potentially realistic realizations of such a scheme  we surveyed in~\cite{Bertolini:2014aia} three setups where the neutrino mass arises at different loop orders, namely, at the tree level via Type-II seesaw~\cite{Schechter:1980gr,Cheng:1980qt,Lazarides:1980nt,Mohapatra:1980yp,Wetterich:1981bx}, at one loop as in~\cite{Babu:2013pma}, and at  two loops as in the Zee-Babu model~\cite{Zee:1985id,Babu:1988ki}, respectively.

 {Regardless of the presence of the large Peccei-Quinn (PQ) scale, in all variants of the scheme discussed in~\cite{Bertolini:2014aia}, a natural and stable electroweak setup is obtained by invoking a decoupling behaviour of the scalar singlet field responsible for the PQ symmetry breaking. All its interactions in the scalar potential (besides the self-interaction)} are scaled down by powers of the electroweak over the PQ scales in such a way that all non-singlet states acquire weak-scale masses. This ``ultraweak'' setting of the singlet scalar interactions is in fact technically stable since the decoupling of the singlet corresponds to an extended Poincar\'e symmetry of the action~\cite{Volkas:1988cm,Bertolini:1990vz}.\footnote{Comments on the role of gravity in such a setting can be found in~\cite{Foot:2013hna}.}

The presence of the invisible axion
 requires at least two Higgs doublets and a complex singlet field. One or more additional scalars are then responsible for the generation of Majorana neutrino masses. 
{The scalar spectra obtained in~\cite{Bertolini:2014aia} are
  naturally compatible with one light SM-like Higgs (a general discussion on the decoupling and alignment limits in a two Higgs doublet context is found in~\cite{Gunion:2002zf}).} On top of that, perturbative naturalness implies that the new scalars should be within the reach of the LHC.
 As a fringe benefit of the extension of the scalar sector the stability of the electroweak vacuum is {expected to be} improved with respect to the SM~\cite{Buttazzo:2013uya,Degrassi:2012ry} (for a recent overview see \cite{DiLuzio:2015iua}).  

{Hence, a} stable renormalizable and realistic extension of the SM {is obtained that not only addresses} the origin of the neutrino masses and mixings {but, at the same time,} connects them to the presence of an invisible axion, a viable dark matter candidate (at variance with the DFSZ model, in the current setup the axion entertains tiny 
couplings to the neutrinos). {These models,} in their minimal realizations, do not exhibit additional sources of CP violation, thus fostering the {dynamical} solution of the strong CP problem via the PQ mechanism~\cite{Peccei:1977hh}. 

In the present paper we investigate the embedding of such a PQ related neutrino mass framework into a classically scale-invariant setup, exploiting the intriguing idea~\cite{Coleman:1973jx} that mass scales {in nature} may originate from quantum effects. Classical scale symmetry is explicitly broken by the renormalization of dimension four interactions. The logarithmic dependence on the mass scales allows for large hierarchies {in terms} of order one ratios of the dimensionless Lagrangian parameters, thus naturally protecting the fundamental mass scale of the theory from any large scale. 

The conjecture that classical scale symmetry may protect the electroweak scale  from large perturbative effects was pioneered in~\cite{Bardeen:1995kv}               
(see~\cite{Hill:2005wg,Hill:2014mqa} for a recent reappraisal). %
{At variance with} the QCD-like strong dynamics, scale invariance is broken by perturbative quantum loops that contribute to the stress-tensor trace anomaly with terms proportional to the beta-functions of the dimensionless couplings  (a pedagogical introduction to scale invariance is found in~\cite{Coleman:1985co}, while its relation with conformal invariance is reviewed in~\cite{Nakayama:2013is}). 

Obtaining a realistic scenario in such a context requires as well an extension of the standard Higgs sector.
The embedding of the DFSZ invisible axion model in a classically scale-invariant setup has been recently discussed in~\cite{Allison:2014hna}. 
As an archetypical implementation of the scale invariant framework to
the neutrino mass models considered in~\cite{Bertolini:2014aia}, we
{focus our analysis on} the  PQ extended Type-II sewsaw
model. The PQ scale, induced via dimensional transmutation,
triggers in turn the electroweak symmetry breaking. The hierarchy between the two scales is set and stabilized by the ultraweak limit {of} the singlet scalar couplings.  Attention is {paid} to the analysis of the scalar spectrum, aiming at a realistic fit of the present LHC data.

The simultaneous presence of the PQ and classical scale symmetries sharply constrain the {scalar potential of the theory}. All but two of the ultraweak couplings are determined by the minimization {conditions} in terms of the other quartic couplings. We show that a natural (i.e., not fine-tuned) and radiatively stable decoupling limit is feasible; in such a case the lightest Higgs boson is fully compatible with the {data} (that require moderate $\tan\beta$ values), while all other physical scalars satisfy the present collider bounds. The needed decoupling and alignment limits of the extra doublet Higgs states are controlled by just one of the two independent ultraweak couplings, while the second one drives both the neutrino mass as well as the decoupling of the scalar triplet states. All this is achieved within a stable and a fully perturbative setup. 

The model exhibits an invisible axion and a very light neutral scalar
that plays the role of a pseudodilaton, both with tiny couplings to
neutrinos that bear no relevance for today's astrophysical and cosmological data (the cosmology of the ultralight pseudodilaton is thoroughly discussed in~\cite{Allison:2014hna}). The smallness of the pseudodilaton mass is due to quantum effects which require a tiny quartic self interaction. This is a characteristic feature of the scale-invariant embedding, at variance with the setups discussed in~\cite{Bertolini:2014aia}, where the strength of the singlet self interaction is unconstrained and a heavy singlet scalar state is allowed.

In summary, {a relatively simple extension of the standard Higgs sector gives rise to} a renormalizable and perturbatively stable scenario where a number of observational and theoretical issues of the SM find a correlated and natural explanation. Were perturbative naturalness a ``fundamental" principle rather than a theorist prejudice, a plethora of new scalar states could be well within the LHC reach.

The study is organized as follows. In the first part of the paper we
introduce the model and study the minimization of the one-loop
effective potential. In the second part we analyze the pattern and phenomenology
of the extended scalar sector and discuss the conditions for vacuum stability. 
Detailed aspects of the analysis are summarized in the appendices.

\section{PQ extended Type-II seesaw}\label{sec:typeIIseesaw}

The field content and the charge assignment of the PQ extended Type-II seesaw model {was worked out in~\cite{Bertolini:2014aia} and it} is displayed for convenience in \Table{fctypeII}. On top of the usual SM field content, the scalar sector includes two Higgs doublets,  
one isospin triplet with {a unit} hypercharge and one complex SM singlet. 
{Since the PQ current is} axial, it is proportional to the difference between the charges of the left- and right-handed 
fermions. Hence, without loss of generality, we may set {the PQ charge of the quark doublets} $X_q = 0$ such that  
the color anomaly of the PQ current turns out to be proportional to $X_u + X_d\neq 0$~\cite{Srednicki:1985xd}.

\subsection{Lagrangian}

The only two sectors which are sensitive to the assignment of the PQ charges are the Yukawa Lagrangian and the scalar potential that we discuss in turn.
The former reads
\begin{multline}
\label{YukawaTII}
- \mathcal{L}^{\rm{TII}}_{Y} = 
Y_u \, \overline{q}_{L} u_{R} H_u 
+ Y_d \, \overline{q}_{L} d_{R} H_d 
+ Y_e \, \overline{\ell}_{L} e_{R} H_d \\ 
+ \tfrac{1}{2} Y_\Delta \, \ell_L^T C i \tau_2 {\Delta}\, \ell_L
+ \rm{h.c.} \, , 
\end{multline}
where the flavour contractions are understood ({$Y_\Delta^{ij} = Y_\Delta^{ji}$}) and
$C$ is the charge conjugation matrix in {the} spinor space.
Borrowing the notation from Ref.~\cite{Bertolini:2014aia}, the classically scale-invariant potential is written as
\begin{align}
\label{TII-scalarpot}
V_0 &=  \lambda_1 \abs{H_u}^4 + \lambda_2 \abs{H_d}^4 + \lambda_{12} \abs{H_u}^2 \abs{H_d}^2 \nonumber \\
 &+ \lambda_{4} \abs{H_u^\dag H_d}^2  + \lambda_{13} \abs{\sigma}^2 \abs{H_u}^2
+ \lambda_{23} \abs{\sigma}^2 \abs{H_d}^2 \nonumber \\
& + \lambda_3  \abs{\sigma}^4  + \Tr ({\Delta}^\dag {\Delta}) \Big[\lambda_{\Delta1} \abs{H_u}^2 + \lambda_{\Delta2} \abs{H_d}^2 \nn \\
& + \lambda_{\Delta3} \abs{\sigma}^2 + \lambda_{\Delta4} \Tr ({\Delta}^\dag {\Delta}) \Big] \nonumber \\
& + \lambda_7 H_u^\dag {\Delta} {\Delta}^\dag H_u 
+ \lambda_8 H_d^\dag {\Delta} {\Delta}^\dag H_d
+ \lambda_9  \Tr ( {\Delta}^\dag {\Delta} )^2 \nn \\
& + \left( \lambda_5 \sigma^2 \tilde{H}^\dag_u H_d + \lambda_6 \sigma H_u^\dag {\Delta}^\dag H_d + \text{h.c.} \right)
\, ,  
\end{align}
where $\tilde{H}_u = i \tau_2 H_u^*$.

\begin{table}[t]
  \centering
  \begin{tabular}{@{} lccccr @{}}
 \hline
  \hspace*{8ex}   & Spin & $SU(3)_C$ & $SU(2)_L$ & $U(1)_Y$ & $U(1)_{PQ}$ \\ 
 \hline
    $q_L$ & $\frac{1}{2}$ & 3  & 2 & $+\frac{1}{6}$ & 0\hspace{1ex} \\ 
    $u_R$ & $\frac{1}{2}$ & 3  & 1 & $+\frac{2}{3}$ & $X_u$ \\ 
    $d_R$ & $\frac{1}{2}$ & 3  & 1 & $-\frac{1}{3}$ & $X_d$ \\ 
    $\ell_L$ & $\frac{1}{2}$ & 1  & 2 & $-\frac{1}{2}$ & $X_\ell$ \\ 
    $e_R$ & $\frac{1}{2}$ & 1  & 1 & $-1$ & $X_e$ \\
    $H_u$ & 0 & 1 & 2 & $-\frac{1}{2}$ & $-X_u$ \\
    $H_d$ & 0 &  1  & 2 & $+\frac{1}{2}$ & $-X_d$ \\
    $\Delta$ & 0 &  1  & 3 & +1 & $X_\Delta$ \\
    $\sigma$ & 0 &  1  & 1 & 0 & $X_\sigma$ \\
 \hline
  \end{tabular}
  \caption{\label{fctypeII} Field content and charge assignment of the PQ extended Type-II seesaw model~\cite{Bertolini:2014aia}. 
   The constraints from the {quark} Yukawa interactions in \eq{YukawaTII} are already taken into account.}
\end{table}

We shall parameterize the complex scalar fields appearing in \eqs{YukawaTII}{TII-scalarpot} as follows
\begin{align}
\label{expHu}
H_u &= 
\left(
\begin{array}{c}
\frac{h^0_u + i \eta^0_u}{\sqrt{2}} \\
h^-_u
\end{array} 
\right) \, , \\ 
\label{expHd}
H_d &= 
\left(
\begin{array}{c}
h^+_d \\
\frac{h^0_d + i \eta^0_d}{\sqrt{2}} 
\end{array} 
\right) \, , \\
\label{expsigma}
\sigma &=
\frac{\sigma^0 + i \eta^0_\sigma}{\sqrt{2}} \, , \\
\label{expdelta}
{\Delta} &\equiv 
\frac{\mathbf{\tau} \cdot {\mathbf\Delta}}{\sqrt{2}}  
= \left(
\begin{array}{cc}
\frac{\delta^+}{\sqrt{2}} & \delta^{++} \\
\frac{\delta^0 + i \eta^0_\delta}{\sqrt{2}} & - \frac{\delta^+}{\sqrt{2}}
\end{array} 
\right)  \, .
\end{align}
In \eq{expdelta}, $\mathbf{\tau} = (\tau_1, \tau_2, \tau_3)$ are the Pauli matrices and 
${\mathbf\Delta} = (\Delta_1, \Delta_2, \Delta_3)$ 
are the $SU(2)_L$ components of the scalar triplet. 
{In what follows,} the vacuum expectation values (VEV) of the neutral components of the scalar multiplets, arising at the quantum level, will be denoted as
\beq
\vev{H_u}=v_u,\ \ \vev{H_d}=v_d,\ \ \vev{\sigma}=V_\sigma,\ \ \vev{\Delta}=v_\Delta \ .
\eeq

{Note that the form of the scalar potential in \eq{TII-scalarpot} is uniquely fixed by} the charge assignment in \Table{fctypeII}~\cite{Bertolini:2014aia}. In particular, terms like $\tilde{H}^\dag_u H_d \Tr ({\Delta}^\dag {\Delta})$ or 
$\tilde{H}^\dag_u {\Delta} {\Delta}^\dag H_d$ 
are not allowed since the QCD anomaly of the PQ current requires $X_u + X_d\neq 0$. 
Moreover, the identity $H_{u,d}^\dag \left( {\Delta}^\dag {\Delta} + {\Delta} {\Delta}^\dag \right) H_{u,d} 
= \abs{H_{u,d}}^2 \Tr ({\Delta}^\dag {\Delta})$ holds, 
so that only two out of three {apparently different} invariants are linearly independent. 

We remind that the interaction terms 
$\lambda_5 \, \sigma^2 \tilde{H}^\dag_u H_d$ and $\lambda_6 \, \sigma H_u^\dag {\Delta}^\dag H_d$ are both needed in order to assign a nonvanishing PQ charge to the singlet $\sigma$ and to generate the neutrino mass. The simultaneous presence of  $\lambda_5$, $\lambda_6$ and $Y_\Delta$ ensures {the} explicit breaking of {the} lepton number.  If any of these couplings is missing, either lepton number is exact and neutrinos are massless ($\lambda_6=0$) or lepton number is spontaneously broken ($\lambda_5=0$) and the vacuum exhibits a {Majoron} together with a Wilczek-Weinberg axion ({in the latter case, discussed in \cite{Bertolini:1990vz}, there exist a charge assignment such that $\sigma$ 
carries two units of lepton number, while being PQ neutral}). 
Both couplings $\lambda_5$ and $\lambda_6$ can be set real by two independent rephasings of the scalar fields.  No spontaneous CP violation arises from such a scalar potential~\cite{Bertolini:2014aia,Haber:2012np}. 

By normalizing 
the PQ charge of the scalar singlet to unity and by imposing its orthogonality to the SM hypercharge (i.e., $X_{u}v_{u}^{2}=X_{d}v_{d}^{2}$) one obtains~\cite{Bertolini:2014aia}  
\begin{align}
&X_u = \frac{2}{\tan^2\beta + 1}\,,& \quad  &X_d = \frac{2 \tan^2\beta}{\tan^2\beta + 1}\, , \nonumber \\
\quad &X_\ell = \frac{\tan^2\beta - 3}{2 (\tan^2\beta + 1)} \,,& \quad&X_e = \frac{5 \tan^2\beta - 3}{2 (\tan^2\beta + 1)}  \, , \nonumber \\ 
\quad &X_\Delta = \frac{3 - \tan^2\beta}{\tan^2\beta + 1}\,,& & 
\label{PQchargesx} 
\end{align} 
where $\tan\beta \equiv v_u / v_d$.

\subsection{PQ scale via dimensional transmutation}

\noindent
A simple but comprehensive discussion of the embedding of the DFSZ invisible axion model~\cite{Zhitnitsky:1980tq,Dine:1981rt} in a classically scale-invariant setup has been recently presented in~\cite{Allison:2014hna}. We shall now analyze the analogous embedding of our Type II invisible axion model, paying attention to the spectrum of the weak-scale scalars, in particular to the conditions for obtaining a 125 GeV SM-like Higgs. In this respect large $\tan\beta$ values are not allowed by the {current} limits on the Higgs down-quark couplings.
The {needed} decoupling of the other scalar eigenstates then further constrains the scalar potential parameter space.

According to the discussion in \cite{Bertolini:2014aia} we assume the technically natural ultraweak limit of the singlet scalar interactions:
\beq
\label{UWlimit}
 \lambda_{i3}, \lambda_5 \sim \mathcal{O}\left(\frac{v^2}{V_\sigma^2}\right), \quad \lambda_6 \sim \mathcal{O}\left(\frac{v_\Delta}{V_\sigma}\right),
\eeq
where $v^2=v_u^2+v_d^2$.
The reference scaling of the singlet couplings is dictated by the requirement that the physical weakly interacting scalars have weak-scale masses.  
It is convenient to introduce the rescaled couplings $c_\lambda$ as
\beq
 \lambda_{i3,5} \equiv c_{i3,5} \frac{v^2}{V_\sigma^2}, \quad \lambda_6 \equiv c_6 \frac{v_\Delta}{V_\sigma}\, .
\label{cnumbers}
\eeq
This setting is at the origin of the hierarchy between the PQ and EW scales and the stability of the Higgs mass, {thus making the setup insensitive to the large PQ scale.}
The limit $\lambda_{i3}, \lambda_5, \lambda_6 \to 0$ 
is associated with the emergence of an additional Poincar\'e symmetry of the action \cite{Volkas:1988cm,Bertolini:1990vz}  (see \cite{Foot:2013hna} for a recent reassessment) 
which makes the ultraweak limit perturbatively stable. It is readily verified that the renormalization of the 
couplings connecting the ``light'' and ``heavy'' sectors is, {as a set,} multiplicative (the relevant beta functions exhibit a fixed point for vanishing couplings, as it is verified from inspection of the one-loop beta coefficients in \app{1looprge}). {Note} that the hierarchy among the ultraweak couplings in \eq{UWlimit} is stable since $\lambda_6^2 \ll \lambda_{i3}$. The couplings $\lambda_5$ and $\lambda_6$ are themselves multiplicatively renormalized since lepton number is restored when one of them is vanishing. 

The stronger scaling pattern of  $\lambda_3$ 
\beq
\label{CWlam3}
\lambda_3 \sim \mathcal{O}\left(\frac{v^4}{V_\sigma^4}\right)
\eeq
is required by the Coleman-Weinberg (CW) mechanism in order to be effective along the singlet direction, so that the PQ scale is obtained by dimensional transmutation~\cite{Coleman:1973jx}. 
This is also a renormalization safe assumption since the limit $\lambda_{i3}, \lambda_5, \lambda_6, \lambda_3\to 0$ is associated with the emergence of a shift symmetry of the noninteracting scalar singlet~\cite{Allison:2014hna}. These considerations hold as long as we neglect gravity. For a brief discussion of gravity induced effects we refer to \cite{Bertolini:2014aia} and references therein.

\subsection{CW potential and the vacuum configuration}

The generation of the electroweak breaking vacuum via the CW mechanism~\cite{Coleman:1973jx} was shown to require a very light Higgs mass, of about 10 GeV~\cite{Coleman:1973jx,Mahanthappa:1979zx}. In this paper,  analogously to the proposal of~\cite{Allison:2014hna}, we provide a realistic setup where quantum correction are responsible for the generation of the PQ scale. This leads, as we will see, to a very light neutral scalar acting as a pseudodilaton, still phenomenologically viable.  The hierarchy between the PQ and {the} electroweak scale is ensured by the technically natural ultraweak limit on the singlet couplings to the other scalar fields~\cite{Volkas:1988cm,Bertolini:1990vz,Bertolini:2014aia}, as stated in \eq{UWlimit}.

To this end the relevant one-loop CW potential in the $\overline{\rm MS}$ scheme can be written as
\begin{equation}
\label{1loopEP}
V_1(\bar\sigma) = \frac{1}{64 \pi^2}  \Tr M^4(\bar\sigma) \left( \log \frac{M^2(\bar\sigma)}{\mu^2} -\frac{3}{2} \right)  \, , 
\end{equation}
where $\bar\sigma$ denotes a spacetime-independent classical field
and the trace is over the tree-level mass matrices in the shifted $\sigma \rightarrow \bar\sigma +\sigma$ theory,
\begin{equation}
\label{M2ij}
M_{ij}^2(\bar\sigma) = \left. \frac{\partial^2 V_0}{\partial \phi_i \partial \phi_j} \right|_{\phi \rightarrow (\bar\sigma,0,...,0)} \, ,
\end{equation} 
where the vector $\phi$ stands for the whole set of {the real fields in the model.} 
In our case the leading contributions to the effective potential in the $\sigma$-field direction read (from now on we drop the bar symbol over $\sigma$)
\begin{align}
	V_{1}= \frac{1}{64 \pi^2} &\left[ 4\, \lambda_{+}^2|\sigma|^4\left(\ln\frac{\lambda_{+}|\sigma|^2}{\mu^2}-\frac{3}{2}\right)\right. + \nonumber \\
	&4\, \lambda_{-}^2|\sigma|^4\left(\ln\frac{\lambda_{-}|\sigma|^2}{\mu^2}-\frac{3}{2}\right) +\nonumber \\
	&\left. 6\, \lambda_{\Delta3}^2 |\sigma|^4\left(\ln\frac{\lambda_{\Delta 3}|\sigma|^2}{\mu^2}-\frac{3}{2}\right)  \right] \, ,
\label{V1}
\end{align}
where one may recognize the complex doublet and triplet contributions.
The singlet quartic interaction is negligible because of \eq{CWlam3}, as detailed in the following. 
The functional masses of the two Higgs doublets in the singlet direction are written in terms of
\begin{equation}
	\lambda_{\pm} = \frac{1}{2}\left(\lambda_{13}+\lambda_{23}\pm\sqrt{(\lambda_{13}-\lambda_{23})^2+4\lambda_5^2}\right) \, .
\end{equation}

We limited the dependence of the functional masses to the $\sigma$ component, assuming that in the other field directions the tree-level quartic couplings dominate the potential. By doing so the perturbative effective potential may develop an imaginary part. This is just a consequence of having effectively chosen a nonconvex point of the one-loop potential in other neutral field directions but the singlet one.  The correct minimum in the $\sigma$ field direction is nevertheless obtained by taking the real part of the effective potential~\cite{Weinberg:1987vp}.
 
According to the original CW approach we minimize the scalar potential along $\sigma$, which is, by construction, the only field direction sensitive to radiative corrections. The {stationarity} equation obtained from the derivative of $V\equiv V_0+V_1$ 
reads
\begin{widetext}
	\begin{align}
		8\pi^2\left.\frac{\partial V_{} }{\partial \sigma}\right|_{\phi=\vev{\phi}} =&\ \left[2\,{\lambda_{{+}}}^{2}\ln   {\frac {\lambda_{{+}}{V_{\sigma}}
				^{2}}{{\mu}^{2}}}  +2\,{\lambda_{{-}}}^{2}\ln   {
			\frac {\lambda_{{-}}{V_{\sigma}}^{2}}{{\mu}^{2}}} +3\,{\lambda_{{
					{\Delta 3}}}}^{2}\ln   {\frac {\lambda_{{{\Delta 3}}}{ V_{\sigma}}^
				{2}}{{\mu}^{2}}}  +32\,{\pi }^{2}\, \lambda_{{3}}-3\,{\lambda_{{{
						\Delta 3}}}}^{2} \right. \nonumber\\ 						
	&  -2\,{\lambda_{{-}}}^{2}  -2\,{\lambda_{{+}}}^{2}
		\Big] V_{\sigma}^3+ 
		16\, \pi^2  \left( \lambda_{{{\Delta 3}}}{{ v_{\Delta}}}^{2}+{\frac {
				\lambda_{{6}}{ v_{\Delta}}\,v_{{d}}v_{{u}}}{V_{\sigma}}}-2\,\lambda_{{5}}v_{{
				d}}v_{{u}}+\lambda_{{13}}{v_{{u}}}^{2}+\lambda_{{23}}{v_{{d}}}^{2}
		\right) V_{\sigma}
		=0 \, .
	\end{align}
\end{widetext}
Of the two terms in the above expression one has a linear leading dependence on $V_{\sigma}$ while the other is cubic. On the other hand, in view of the ultraweak limit in \eq{UWlimit}, both terms are of the same order and must be equally considered.
We see that a stationary point exists either when $V_{\sigma}=0$ or $V_{\sigma}$ arbitrary and
\begin{align}
	\lambda_{{3}}= \frac{\lambda_3^\prime v^2}{2V_{\sigma}^2} -{\frac {1}{16 \pi^2}}\,  &\left[\,{\lambda_+^2}{
		} \left( \ln {
		\frac {\lambda_{{+}}{V_{\sigma}}^{2}}{{\mu}^{2}}}  -1\right)\right.    \nonumber  \\
&	+{\lambda_-^2} \left(\ln   {
		\frac {\lambda_{{-}}{V_{\sigma}}^{2}}{{\mu}^{2}}}  -1\right) \nonumber \\
&		+  \left.\frac{3}{2}\,{\lambda_{{  \Delta 3}}^2} \left(\ln  {
		\frac {\lambda_{{\Delta 3 }}{V_{\sigma}}^{2}}{{\mu}^{2}}}  -1 \right)\right] 
\label{lambda3}
\end{align}
where we defined
\begin{align}
	\lambda_3^\prime v^2 \equiv &\ 2 \lambda_5v_uv_d-\lambda_{13}v_u^2-\lambda_{23}v_d^2-\lambda_{\Delta 3}v_{\Delta}^2  
	\nonumber\\
&	-\lambda_6v_uv_dv_{\Delta}/V_{\sigma} \, ,
\label{veps}
\end{align}
with $\lambda_3^\prime = O(v^2/V_\sigma^2)$.
As mentioned above, since the functional masses in \eq{V1} are taken at a non convex point of the perturbative effective potential ($v_{u,d}=0$) an imaginary part may develop~\cite{Weinberg:1987vp}. Accordingly, the real part of \eq{lambda3} is understood.

We have traded the dimensionless parameter $\lambda_3$ for $\vev{\sigma}/\mu$.
Obviously, for the CW mechanism to work an (almost) flat field direction is needed in the scalar potential. In the case at hand the singlet quartic coupling 
$\lambda_3$ is required to be of the order of the square of the other ultraweak couplings and loop suppressed (this justifies its {omission from} the one-loop effective potential). 

By plugging the above expression for $\lambda_3$ {into} the  potential one finally finds 
\begin{widetext}
	\begin{align}
		V_{\text{eff}}=&\ (V_0-\lambda_3|\sigma|^4) +\frac{\lambda_3^\prime v^2}{2 V_{\sigma}^2}|\sigma|^4
		+{\frac {1}{16\pi^2}}\left[\,{\lambda_{{+}}}^{2}  +\,{\lambda_{{-}}}^{2}+\tfrac{3}{2}\,{\lambda_{{{ \Delta 3}}}}^{2}
		\right]  \left( \ln  {\frac {{|\sigma|^2}}{V_{\sigma}}}  -1/2 \right)	|\sigma|^{4} \nonumber \\
		=&\ (V_0-\lambda_3|\sigma|^4) +\frac{\lambda_3^\prime v^2}{2 V_{\sigma}^2}|\sigma|^4
		+{\frac {\beta_{\lambda_{3}}}{2}} \left( \ln  {\frac {{|\sigma|^2}}{{V_\sigma}^{2}}}  -1/2 \right)|\sigma|^4 \, ,
\label{Veff}		
	\end{align}  
\end{widetext}
where $\beta_{\lambda_{3}}$ is the one-loop beta function of the singlet quartic coupling, as given in
\app{1looprge}.

By comparing our result with the renormalization group (RG) based
discussion of \cite{Allison:2014zya,Allison:2014hna} one sees that a
term analogous to $\lambda_3^\prime$ is missing in the effective
potential. The difference amounts to a redefinition of $V_\sigma$ (or
a shift on the {singlet} quartic coupling) which does not bear any physical consequences.  

The effective potential in \eq{Veff} does not explicitly depend on the renormalization scale, since the explicit $\mu$ dependence of $\lambda_3$ in \eq{lambda3} precisely cancels the explicit  $\mu $ dependence of the one-loop effective potential. It is worth remarking that the coefficient of the quartic term for the singlet coincides precisely with the beta function of $\lambda_3$ (which includes a negligible contribution of the $\sigma$ quartic self interaction, proportional to $\lambda_3^2$).

The minimization proceeds with the derivation of the stationarity equations in the remaining field directions, where quantum corrections are safely neglected:

\begin{align}
&\frac{\partial V}{\partial v_u}=2\,{{  v_{\Delta}}}^{2}\lambda_{{{  \Delta 1}}}v_{{u}}+2\,{  v_{\Delta}}\,\lambda_{{
		6}}v_{{d}}V_{\sigma}+4\,\lambda_{{1}}{v_{{u}}}^{3}-2\,\lambda_{{5}}v_{{d}
}{V_{\sigma}}^{2} \nonumber \\ &+2\,\lambda_{{12}}{v_{{d}}}^{2}v_{{u}}+2\,\lambda_{{13}}
{V_{\sigma}}^{2}v_{{u}}=0 \label{eqnvu} \, , \\
%
&\frac{\partial V}{\partial v_d}=2\,{{  v_{\Delta}}}^{2}\lambda_{{8}}v_{{d}}+2\,{{  v_{\Delta}}}^{2}\lambda_{{{
			\Delta 2}}}v_{{d}}+2\,{ \lambda_{{6}} v_{\Delta}}\,V_{\sigma}v_{{u}}+4\,\lambda_
{{2}}{v_{{d}}}^{3} \nonumber \\
&-2\,\lambda_{{5}}{V_{\sigma}}^{2}v_{{u}}+2\,\lambda_{{
		12}}v_{{d}}{v_{{u}}}^{2}+2\,\lambda_{{23}}v_{{d}}{V_{\sigma}}^{2}=0 \label{eqnvd}  \, , \\
%
&\frac{\partial V}{\partial v_{\Delta}}= 4\,{{  v_{\Delta}}}^{3}\lambda_{{9}}+4\,{{  v_{\Delta}}}^{3}\lambda_{{{  \Delta 4}}}
+2\,{  v_{\Delta}}\,\lambda_{{8}}{v_{{d}}}^{2} +2\,{  v_{\Delta}}\,\lambda_{{{
			\Delta 1}}}{v_{{u}}}^{2}\nonumber \\
&+2\,{  v_{\Delta}}\,\lambda_{{{  \Delta 2}}}{v_{{d}}}^{2}
+2\,{  v_{\Delta}}\,\lambda_{{{  \Delta 3}}}{V_{\sigma}}^{2}+2\,\lambda_{{6}}v_{{d
	}}V_{\sigma}v_{{u}}
	=0 \label{eqnvs}  \, .
\end{align}
{By taking into account \eq{UWlimit} and that electroweak measurements bound $v_\Delta /v$ to be less than a percent, \eq{veps} can be cast into the form}
\beq
\lambda_3^\prime v^2 \approx \frac{2}{V_{\sigma}^2}(\lambda_1v_u^4+\lambda_2 v_d^4+\lambda_{12}v_u^2v_d^2)\, ,
\label{veps2}
\eeq
{which holds up to $O(v^2_\Delta /v^2)$ corrections.}

It is worth remarking that due to the ultraweak size of the singlet couplings the relevant scale for the minimization of the effective potential is not the PQ scale, but rather the electroweak scale. The latter in fact minimizes the logarithmic terms in \eq{V1}, and therefore minimizes higher-order corrections to the vacuum. We shall therefore consider the stationarity equations as constraints on the couplings evaluated at the weak scale.

\subsection{Scalar spectrum}
\label{scalarspectrumTII}

The scalar spectrum of the model is worked out in detail in \app{scalarspectrum}. 
Considering the hierarchy among the VEVs ($v_\Delta\ll v_{u,d}\ll
V_\sigma$) and the ultraweak limit we may safely neglect {at the
percent level accuracy} the mixings of the triplet with the other scalars and {perform an analytic diagonalization} of the doublet-singlet mass matrix.

{The results can be cast in fairly simple form. Of four neutral CP-even scalars, two are} made mostly of the $h_u^0$, $h_d^0$ doublet fields (see \eqs{tildeh}{tildeH} in \app{scalarspectrum}). 
The mass eigenvalues are given in \eq{hHtildemasses}. {In view of the results of the phenomenological discussion to follow,} it is convenient to give their expression in the limit of large $c_5 \gg \lambda_{1,2,12}$, namely 
 	\begin{align}
 m_h^2 \approx &\ 2 \left(\lambda_1v_u^2+\lambda_2v_d^2\right) 
            + 2 \left(\lambda_1v_u^2-\lambda_2v_d^2\right) \frac{\tan\beta-\cot\beta}{\tan\beta+\cot\beta} \nn\\
            &+ \lambda_{12} v^2 \sin^2 2\beta
 	\label{hmasslargec5}
 	\end{align}
and 
 	\beq
 m_H^2 \approx c_5 v^2 (\tan\beta+\cot\beta) \, ,
	\label{Hmasslargec5}
 	\eeq
respectively ($\tan\beta=v_u/v_d$). \eqs{hmasslargec5}{Hmasslargec5} exhibit the large $\tan\beta$ ($\cot\beta$) mass dependence as well.
The lighter state $h$ will be identified with the SM-like Higgs. The mixing between the two neutral scalars is parametrized by the angle $\alpha$ 
(defined in \eq{alphamix} of \app{scalarspectrum}). For $\cos(\alpha-\beta)\to 0$  the couplings of the lightest eigenstate overlap with those of the SM Higgs (alignment limit~\cite{Gunion:2002zf}). {The other neutral scalar ($H$)} can be made parametrically heavier, {with mass} in the TeV range, by increasing $c_5$. Correspondingly, $\cos(\alpha-\beta)$ decreases, as we shall shortly detail. 

The remaining two neutral scalars coincide with high precision with the $\sigma^0$ and  $\delta^0$ fields, respectively. Their masses are given by
\begin{align}\label{dilatonmass}
&m_\sigma^2 \  = 2 \beta_{\lambda_3} V_\sigma^2\, ,\\
&m_{\Delta_S}^2 = -c_6 v_u v_d\, ,
\end{align}
where $c_6<0$. The coefficients $c_\lambda$, defined in \eq{cnumbers}, are {${\cal O}(1)$} parameters that gauge the degree of naturalness of the model. Since the mass eigenstates {scale roughly with their square root}, deviations within a factor {of 10} from unity are to be considered natural~\cite{Bertolini:2014aia}. The very light (mainly) singlet eigenstate can be identified with a pseudodilaton~\cite{Allison:2014zya,Allison:2014hna,Bellazzini:2013fga,Coradeschi:2013gda}. The ultraweak character of its couplings underlies a shift symmetry that makes in the limit the field $\sigma$ formally analogous to a dilaton, as arising from spontaneous breaking of the scale invariance.
{Due to} the ultraweak setup in \eq{UWlimit} its mass is bound to be below the MeV scale. Since the CW mechanism is effective only in the $\sigma$ direction, it is {also} the only scalar state whose mass is determined by quantum effects, as a consequence of the explicit breaking of {the} scale invariance. 

The pseudoscalar spectrum consists of the {neutral} Goldstone
boson (GB) ``eaten'' by $Z$ ({spanned predominantly on $\eta^0_u$
  and $\eta^0_d$}), the invisible axion (mostly $\eta^0_\sigma$),
which receives a tiny mass from QCD instantons, and the fields $A$
({mainly a combination of  $\eta^0_u$ and $\eta^0_d$ orthogonal to the GB above}) and {$\Delta_{A}$} (mostly $\eta^0_\delta$) with masses
\begin{align}
&m_{A}^2  = c_5 v^2 (\tan\beta+\cot\beta) \, ,\\
&m_{{\Delta_A}}^2 = -c_6 v_u v_d \, ,
\end{align}
where $c_5>0$. The leading order (LO) equality of $\Delta_S$ and $\Delta_A$ masses is a consequence of the {neglect of} the triplet mixings with the other scalars 
(see \app{scalarspectrum}).

Among the singly-charged scalars one {of the mass eigenstates is the charged GB ``eaten'' by $W^{\pm}$ and there are} two massive states (mostly $h_d^+$ and $\delta^+$, respectively):
\begin{align}
m_{H^{+}}^2 &= \lambda_4v^2+ c_5 v^2 (\tan\beta+\cot\beta),\\
m_{{\Delta^+}}^2& = \tfrac{1}{2}(\lambda_{{7}}{v_{{u}}}^{2}-\lambda_{{8}}{v_{{d}}
     	}^{2}) -\,c_{{6}}v_{{d}}v_{{u}}.
\end{align}

Finally, the doubly-charged component of the triplet field acquires the mass 
\beq
m_{\Delta^{++}}^2 = \lambda_{{7}}{v_{{u}}}^{2}-\lambda_{{8}}{v_{{d}}
     	}^{2} -\,c_{{6}}v_{{d}}v_{{u}}.
\eeq
Notice the $\tfrac{1}{2}(\lambda_{{7}}{v_{{u}}}^{2}-\lambda_{{8}}{v_{{d}}}^{2})$ mass isosplitting among the components of the scalar triplet. 

While the {masses} of the (mainly) doublet and triplet states are
driven by the tree-level part of the potential, {the weak-scale}
pseudodilaton mass (the ultraweak setup is assumed) is genuinely obtained at one loop. As such \eq{dilatonmass} should be renormalized down to its characteristic scale ($\ll$ MeV), and {finite momentum corrections should be included}. On the other hand, the pseudodilaton couples to matter and gauge {fields} only through its mixings with the doublet and triplet scalars which are suppressed by the PQ scale and, {thus,}  the running of the pseudodilaton mass is of a higher order.  Analogously, finite momentum dependent one-loop corrections are suppressed by the mass of the heavier particles in the loop, and for the scope of the present analysis they can be neglected as well.

The physics of the invisible axion is analogous to that of the DFSZ model
with the addition of tiny couplings to the neutrinos that, {however,} do not bear observable cosmological or astrophysical implications~\cite{Bertolini:2014aia}. A short discussion on the light pseudodilaton phenomenology is deferred to~\sect{sec:dilatonpheno}.

\begin{figure}[t]
\begin{center}
\includegraphics[width=.5\columnwidth]{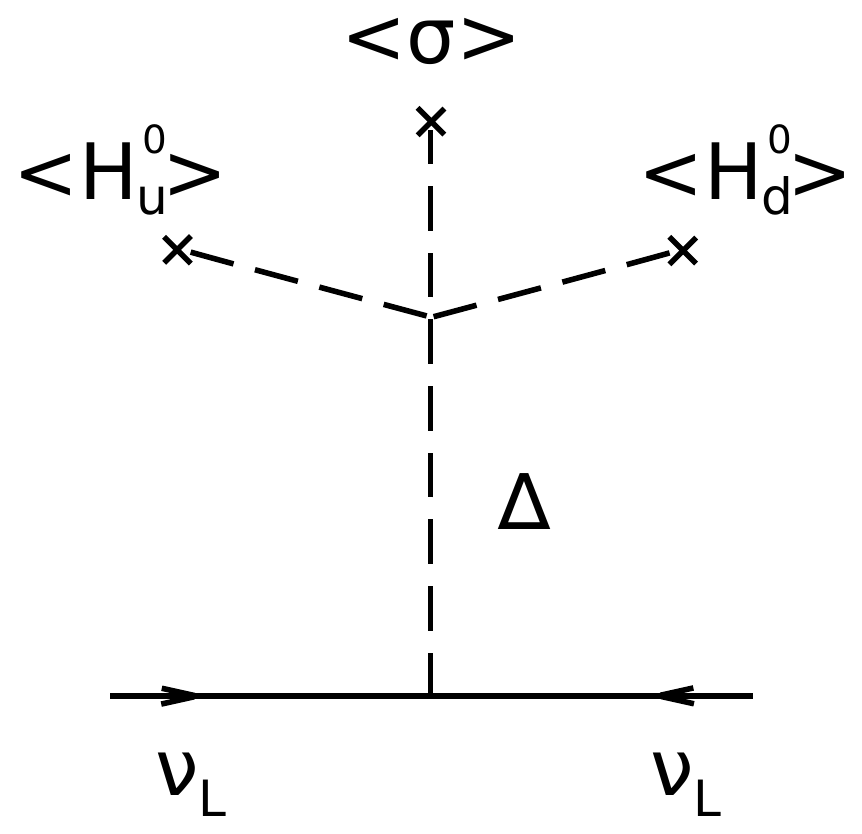}
\caption{The ``hug'' diagram, responsible for the neutrino mass in the PQ extended Type-II seesaw model.}
\label{Fig:axseesawII}
\end{center}
\end{figure}

\subsection{Neutrino masses}

In the PQ extended Type II seesaw model~\cite{Bertolini:2014aia}, the neutrino masses are generated through the tree-level diagram in \fig{Fig:axseesawII}. The {corresponding (symmetric)} mass matrix is readily obtained 
from the Yukawa Lagrangian in \eq{YukawaTII} as
\begin{equation}
\label{neutrinomasses}
M_\nu = Y_\Delta v_\Delta \approx -\frac{Y_\Delta \lambda_6 V_\sigma v_u v_d}{M^2_\Delta} \, ,
\end{equation}
{where $M_\Delta$ is the common mass of the neutral triplet components.}
Consequently, the bound on the heaviest neutrino $m_{\nu_3} \lesssim 1$ eV translates into the constraint 
\begin{equation}
|\lambda_6| Y_\Delta \lesssim 10^{-18} \left( \frac{10^9 \ \rm{GeV}}{V_\sigma} \right)  \, .
\end{equation}
The smallness of the {absolute neutrino mass scale} may have different {origins} and the naturalness of the setup plays a relevant role. 
{The fact that $M_\Delta$ is by construction not far from the
  electroweak scale sharply constrains the size of the triplet Yukawa
  couplings $Y_\Delta^{ij}$ in the tree-level lepton flavor violation
  processes~\cite{Bertolini:2014aia}. Complementary to that,
the smallness of the ultraweak coupling $\lambda_6\approx v_\Delta/V_\sigma$ is a crucial factor for
the required neutrino mass suppression, that can be
obtained even for somewhat large $Y_\Delta$, depending on the actual
size of $v_\Delta$. }

\section{Higgs phenomenology and vacuum stability}     
\label{vacuumstability}

{In the technically natural ultraweak limit of \eq{UWlimit} a number of new scalar states may possibly fall within the reach of the present and near future} collider searches.

In this section we shall discuss the conditions {under} which the scalar spectrum is phenomenologically viable and the vacuum structure of the model is stable.  
A {necessary prerequisite to that} is the analysis of the constraints on the scalar couplings coming from {the measured value of the Higgs mass  ($125.09 \pm 0.21 \pm 0.11$ GeV~\cite{Aad:2015zhl})} and the allowed deviations {of} its couplings to matter and gauge fields 
(for a recent study of {the Higgs LHC physics and the} electroweak precision observables in the DFSZ ultraweak setup see~\cite{Espriu:2015mfa}).
We shall give a benchmark setting of the model parameters that satisfies all the present LHC constraints with the heavier scalars possibly below the TeV scale, and thus accessible {to the collider searches}. We also include a brief phenomenological overview of the light pseudodilaton phenomenology.

\subsection{Experimental constraints}

The Yukawa couplings of the lightest {neutral scalar} eigenstate $h$ can be parametrized at the LO in $\cos({\beta-\alpha})\ll 1$ as~\cite{Gunion:2002zf}
\begin{align}
	\mathcal{L}_Y \approx &\  \frac{1-\tan\beta \cos({\beta-\alpha})}{v} \bar{D}M_DD\, h \nonumber \\ 
	&  +\frac{1+\cot\beta \cos({\beta-\alpha})}{v} \bar{U}M_UU\, h \, ,
\end{align}
where $\alpha$ is the mixing angle between the light and heavy eigenstates (see \app{scalarspectrum}).

{Analogously}, the $h$ couplings with the gauge bosons are conveniently written as
\begin{align}
	\frac{g_{hVV}}{{g_{hVV_{SM}}}}\approx 1-\frac{\cos^2(\beta-\alpha)}{2} \, .
\end{align}

\begin{figure}[t]
     	\includegraphics[width=0.5\textwidth]{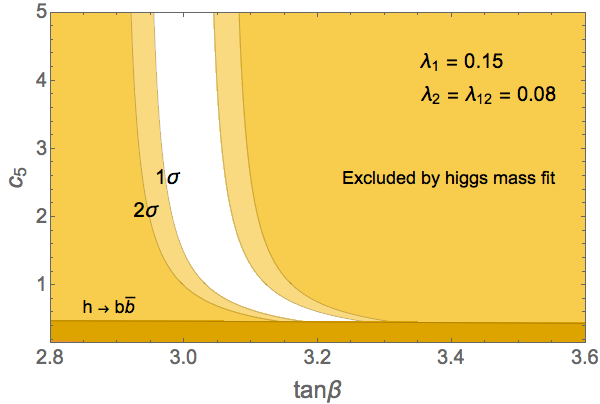} 
     	\caption{\label{constraints1} Allowed region at 1$\sigma$ and $2\sigma$ in the $(c_5,\tan\beta)$ plane from the constraints coming from the Higgs mass and the decays $h\rightarrow b\bar{b}$,   $h\rightarrow t\bar{t}$ and $h\rightarrow VV$.}
     \end{figure}

{Based on the results in \app{scalarspectrum},} \fig{constraints1} shows the constraints {on $c_5$ and $\tan\beta$ coming} from the present fit of the Higgs mass and couplings~\cite{Corbett:2015ksa} 
(with benchmark values of the relevant quartic couplings).  
 Only moderate values of $\tan \beta$ are allowed.
 {The strong constraint on $\tan\beta$ is a consequence of the
   specific} pattern of the scalar couplings and can be readily
 inferred from \eq{hmasslargec5}. In the large $c_5$ limit the leading
 $c_5$ contributions cancel in the lightest scalar {mass}
 eigenvalue (while they sum up in the heavier {one}). The assumed
 perturbativity of the scalar couplings up to the Planck scale
 together with the constraints from the Higgs mass lead to the
 {typical} $10^{-1}$ scale for the relevant doublet couplings. For
 doublet couplings of similar size a double solution above and below $\tan\beta=1$ appears, as expected.  A value of $\tan\beta$ near unity is allowed for $\lambda_1 = \lambda_2 = \lambda_{12}\approx 0.17$ {which, however, is too large} for perturbativity.
On the other hand, when considering lower values, a hierarchy between
$\lambda_1$ and the other  doublet couplings is imposed {by the
  requirement of} vacuum stability ($\lambda_1$ is affected by a large top quark
negative renormalization and it is bound to larger values). This, {altogether,} selects the $\tan\beta > 1$ solution. For decreasing $c_5 < 1$ a solution is maintained by increasing $\tan \beta$. These features are apparent in \fig{constraints1} for {typical values of the} doublet couplings. {The specific pattern there is a consequence of the hierarchy between $\lambda_1$ and the other doublet couplings which is needed} to reconcile perturbativity with vacuum stability. 
{The ``lower part'' of the plot (i.e., $c_{5}\lesssim 0.5$) is cut
  out by the constraints from the $h\rightarrow b\bar{b}$ data that
  still allow for about 30\% deviation from the SM
  coupling~\cite{Corbett:2015ksa}. In combination, a rather sharp
  constraint on $\tan\beta$ emerges.}

      \begin{figure}[t]
     	\includegraphics[width=0.485\textwidth]{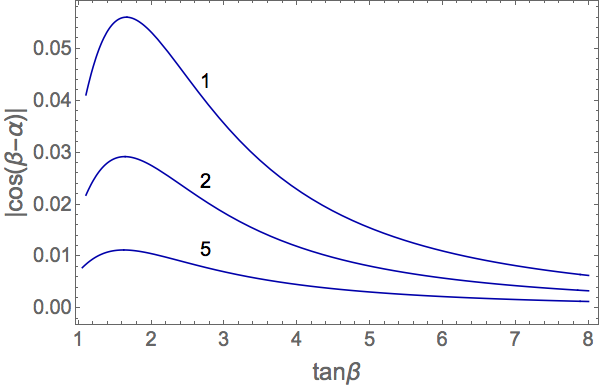} 
     	\caption{\label{cosbetamalpha} Dependence of $\cos(\beta-\alpha)$ on $\tan\beta$ for different values of $c_5$.}
     \end{figure}

      \begin{figure}
     	\includegraphics[width=0.5\textwidth]{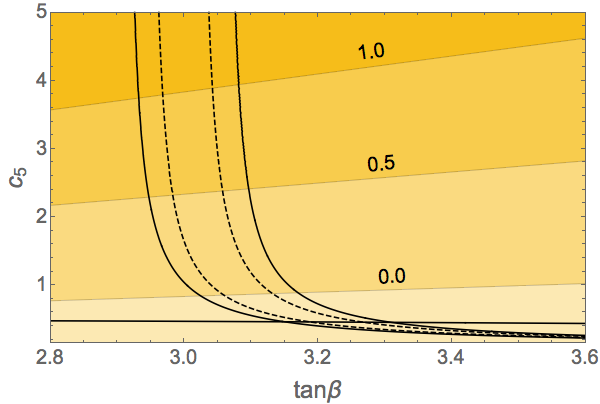} 
     	\includegraphics[width=0.5\textwidth]{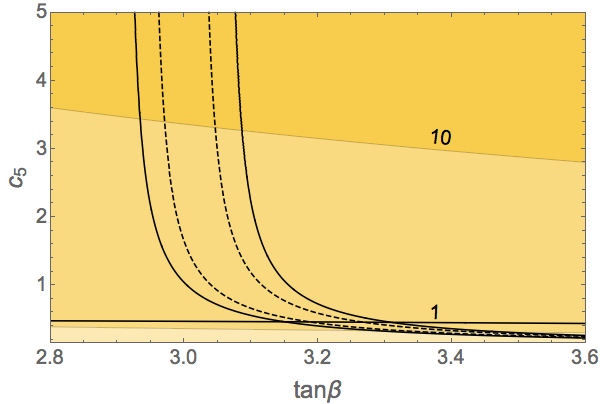} 
     	\caption{\label{contour2} Contour lines showing the values of the rescaled coupling $c_{13}$ in the $(c_5,\tan\beta)$ plane (top panel). Analogous contours for $c_{23}$ (bottom panel). The dashed lines show the allowed region from~\fig{constraints1}.  }
     \end{figure}

A phenomenologically acceptable mass gap between the lightest Higgs and the other physical doublet eigenstates can then be obtained, even for moderate values of $\tan\beta$ and perturbative quartic couplings, by raising $c_5$, as we shall shortly detail (no issue of perturbativity arises since $\lambda_5$ is an ultraweak coupling). In the large $c_5$ limit the doublet mixing angle 
$\alpha$ is simply related to $\tan\beta$. From \eq{alphamix} one obtains
\beq
\tan 2\alpha \approx \frac{2}{\cot\beta-\tan\beta}
\label{alphalargec5} \, .
\eeq  
According to the doublet eigenstates and mixings defined in \app{scalarspectrum},  $\sin 2\alpha<0$ in the limit. This leads to $\cos(\beta-\alpha)\approx0$, which corresponds to the alignment with the SM Higgs couplings discussed in~\cite{Gunion:2002zf}.
 We recall that, in the broken PQ phase, the $\lambda_5$ quartic coupling induces a mixing between the two Higgs doublets. The pattern of the scalar doublet spectrum for large $c_5$ {then} corresponds to the decouping limit for large $m_{12}^2$ in Ref.~\cite{Gunion:2002zf}. 
 
 The detailed dependence of $\cos(\beta-\alpha)$ on $\tan\beta$ is shown in
      \fig{cosbetamalpha} for the benchmark values of the couplings in \fig{constraints1} and {few} typical values of $c_5$. This allows {one to readily estimate} the size of the deviations of the light Higgs couplings to quarks and gauge bosons {from their SM values}. In particular, we understand why the present experimental uncertainties do not strongly affect the model even for moderate values of $\tan\beta$:  
     {even considering the largest allowed value of $\tan\beta$ in \fig{constraints1},  the present experimental constraints} on the $hb\bar b$ couplings lead to $|\cos(\beta-\alpha)| < 0.09$, well above all model values  for $c_5 > 1$, cf. Fig.~\ref{cosbetamalpha}. When the constraints on the Higgs couplings become stronger the mixing angle $\alpha$ must be further tuned toward its effective decoupling value $\beta -\pi/2$ by {means of} a larger $c_5$, with a corresponding increase {in} the mass of the heavier ``doublet states'' and a progressive loss of naturalness of the setup.

     It is remarkable that the deviations of the couplings of the {light Higgs boson} to the top quark and to the gauge bosons are always very small (below 1\%  in the allowed range of $\tan\beta$ for $c_5>1$). This justifies the assumption of the {usual} SM Higgs production {rates
used} in our numerical analysis.

     In \fig{contour2} we show the contour plots for the values of the ultraweak couplings   $\lambda_{13}$ and $\lambda_{23}$ in the $(c_5,\tan\beta)$ plane, in the same area of the parameter space shown in \fig{constraints1}. 
       We see that $c_{13}$ may be negative for small $c_5$, while $c_{23}>1$ in the allowed region. The fact that  $\lambda_{13}$ may be negative at the weak scale is not an issue for the vacuum stability since, as we shall see, the positivity of the relevant boundedness condition, which involves other quartic couplings, is always satisfied.

        \begin{table}
      	\begin{center}
      		\begin{tabular}{lr} 
      			\hline
      		 Quartic coupling \hspace{5em} & Electroweak-scale value \\ [0.5ex] 
      			\hline
      			$\lambda_{1}$& 0.15   \\ 
      			
      			$\lambda_{2}=\lambda_{12}=\lambda_{4}$ & 0.08   \\ 
      			
      			$\lambda_{7}=\lambda_{8}=\lambda_{9}$ & 0.08 \\
      			
     		$\lambda_{\Delta 1}=\lambda_{\Delta 2}=\lambda_{\Delta 4 }$ & 0.08 \\
      						
      				$\lambda_5/c_5$ &  $3.0 \times 10^{-14}\left(\tfrac{10^{9}\ \rm{GeV}}{V_\sigma}\right)^2$\\
      				
      				$\lambda_6/c_6$ &  $1.0 \times 10^{-9} \left(\tfrac{v_\Delta }{\rm{GeV}}\right)\left(\tfrac{10^{9}\ \rm{GeV}}{V_\sigma}\right)$\\
      					
       			$\lambda_{13}(*)$& $1.3 \times 10^{-15}$   \\ 
      			 
      			$\lambda_{23}(*)$ &  $9.1 \times 10^{-14}$\\
      			
      			$\lambda_{\Delta 3}(*)$ &  $6.2 \times 10^{-15}$\\
      			
      				$\lambda_{ 3}(*)$ &  $1.1 \times 10^{-28}$\\
       				\hline
      		\end{tabular}
      		\caption{\label{benchmarkL}  Benchmark values of quartic couplings at the electroweak scale $v\approx 174$ GeV. The starred couplings are related by the vacuum to the other ones and their reference values are given for $\tan\beta = 3.1$, $c_5=-c_6=1$ and $V_{\sigma}=10^{9}$~GeV ($v_\Delta< 1 $~GeV). As discussed in the text the tiny scale of the ultraweak couplings is protected by symmetry and it is therefore technically natural.}
      	\end{center}
      \end{table}

In \Table{benchmarkL} the chosen benchmark values of the independent scalar couplings are reported. The starred ultraweak couplings are related by the stationarity equations to the other couplings. They are given for a typical value of $\tan\beta$ and the reference values of $\lambda_{5,6}$, namely $c_{5,6}=1$ which set the scalar spectrum at the weak scale (and the size of the other ultraweak couplings). The smaller value of $\lambda_{13}$ is related to the crossing from positive to negative values near the benchmark point (see \fig{contour2}). The self interaction $\lambda_3$ scales with the square of the other ultraweak couplings, as required by the CW mechanism.

   In \Table{doublet} and \Table{triplet} typical mass  {spectra of the
   scalar fields are displayed.}
   The mass scale of the exotic doublet and triplet states is controlled by $c_5$ and $c_6$, respectively. The large {value}
of $c_6$ which drives the triplet masses does not affect neither the stability of the $\lambda_6$ coupling (which, alike $\lambda_5$, is multiplicatively renormalized), nor perturbativity since $\lambda_6$ is ultraweak. Rather, as for $c_5$, it is a {measure} of the naturalness of the setup. Roughly {speaking}, the degree of fine tuning related to the stability of the lightest Higgs mass against radiative corrections induced by the scalar triplet states is proportional to the square root of $|c_6|$.    
    
    This analysis allows us to conclude that the model can accommodate all the present  experimental constraints~\cite{ATLAS:2013zla,Khachatryan:2014jya,ATLAS:2014kca}, while maintaining, as we will shortly detail, vacuum stability and absence of Landau poles up to the {Planck} scale.
 {Indeed,}       
 as already mentioned, among {all} the Higgs doublet couplings, $\lambda_1$ takes the largest value in order to avoid the vacuum instability due to the large renormalization effect induced by the top quark. The smaller values chosen for the other couplings
{then} ensure the absence of Landau poles below the Planck scale.

     \begin{table}[t]
     	\begin{center}
     		\begin{tabular}{l r rrr r} 
     			\hline
     			$\tan \beta$ & $c_5$  &\hspace{1em} $m_h[\text{GeV}]$ & $m_H [\text{GeV}]$& $m_A[\text{GeV}]$ &$m_{H^+}[\text{GeV}]$ \\ [0.5ex] 
     			\hline
     			3.1 & 1.0 & 125 &  324 & 322 & 326 \\ 
     			
     			3.0 & 2.0 & 125 &  451 & 449 & 452 \\ 
     			
     			3.0 & 5.0 & 125 &  711 & 710 & 712 \\ 
      			\hline
     		\end{tabular}
     		\caption{\label{doublet}  Typical values of the doublet scalar masses for the values of the quartic couplings in \Table{benchmarkL} when varying $\tan\beta$ and $c_5$ within the allowed region in \fig{constraints1}.  }
     	\end{center}
     \end{table}
       \begin{table}[t]
       	\begin{center}
       		\begin{tabular}{l r  ccc r} 
       			\hline
       			$\tan \beta$ & $-c_6$  &\hspace{1em} $m_{\Delta^0}[\text{GeV}]$ & $m_{\Delta^+} [\text{GeV}]$& $m_{\Delta^{++}}[\text{GeV}]$ &$m_{\sigma}[\text{GeV}]$ \\ [0.5ex] 
       			\hline
       			         & 25  & 476 & 477 & 478 &  $   0.5 \times 10^{-4} $\\ 
       			
       			3.0    & 50 & 674 &  674 &  675  & $  0.9 \times 10^{-4} $ \\ 
       			
       			         & 75  & 825 &  826 & 826 & $   1.3 \times 10^{-4} $\\ 
       			\hline
       		\end{tabular}
       		\caption{\label{triplet}  Typical triplet and pseudodilaton masses for the benchmark values of the quartic couplings in \Table{benchmarkL}. The scalar masses are given for reference $\tan\beta$ and $c_6$ values that accommodate the present collider limits on the doubly-charged triplet component. The dependence of the pseudodilaton mass on $c_5$, in the interval 1 to 5, ranges from a factor of 2 to a factor of 1.1 as  $c_6$ increases.}
       	\end{center}
       \end{table}

\begin{figure}[t]
     	\includegraphics[width=0.5\textwidth]{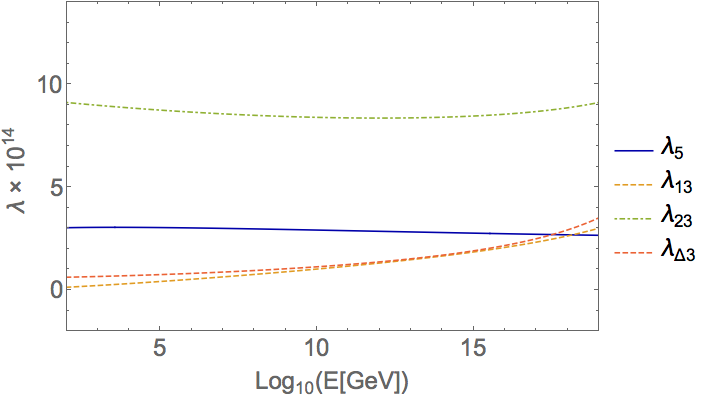} 
     	\caption{\label{lambdaUW} One-loop running of ultraweak couplings for the weak-scale benchmark values of \Table{benchmarkL}.  }
     \end{figure}

A few comments on the pattern of the scalar couplings are in order.
Among the ultraweak couplings that, {as a set,} renormalize
multiplicatively,  a large hierarchy may appear with respect to $\lambda_6$ (it actually depends on the value of the triplet VEV which is bound to be smaller than about 1 GeV). On the other hand, this hierarchy does not destabilize the ultraweak couplings as long as the size of $\lambda_6$ is of the order of the square root of the {other ultraweak couplings} or smaller. This is a consequence of the fact that the $\lambda_6$ interaction involves four different scalar multiplets and therefore affects the renormalization of the other couplings quadratically at one loop (see \fig{lambdaUW}). In terms of the rescaled coupling $c_6$ such a condition can be conservatively written as
\beq
|c_6| < \frac{v}{v_\Delta}
\label{c6bound} \, ,
\eeq
independently on the PQ scale, $V_\sigma$.
Depending on the actual value of $v_\Delta < 1$ GeV, stability of the ultraweak setup is maintained even in the presence of very large values of $c_6$ that, {indeed, may be needed by the} heavy triplet states.

Finally, $\lambda_5$ and $\lambda_6$ are individually multiplicatively
renormalized since lepton number is restored when one of them
vanishes. In the limit $c_5\to 0$ {there is a PQ charge assignment
  that leaves the singlet scalar carrying only} the lepton number. We
thus recover at the tree level two massless pseudoscalar states: an
invisible {Majoron} and a weak-scale axion. In the limit $c_6\to 0$
lepton number is restored and {it remains unbroken as there is no
  induced triplet VEV}. From the inspection of the triplet mass
spectrum we observe the vanishing of the mass of the  neutral triplet
components, while for $\lambda_{7,8}\to 0$ all triplet-dominated
fields turn out to be  massless as well. These tree-level results are related to the neglect of the triplet mixings, and they can be generally understood in terms of accidental shift symmetries of the scalar potential for vanishing triplet couplings. 
All these features are explicitly verified by the inspection of the mass spectrum and the one-loop beta coefficients reported in \app{1looprge}.

The tiny value of  the quartic singlet coupling is also preserved by renormalization since only {squares of} the ultraweak couplings appear in its beta function. This is again understood in terms of symmetries since, for vanishing interactions of the singlet with the other fields, a further shift symmetry arises when the quartic self interaction vanishes as well. All that can be explicitly seen by inspection of the one-loop beta coefficients given in \app{1looprge}. We can therefore conclude that the pattern of the benchmark values given in \Table{benchmarkL} is technically natural, and  leads (within the present experimental limits on the scalar spectrum) to a natural and stable model setup.  

 From \Table{doublet} and \Table{triplet} we see that the {heavy part of the scalar spectrum of the model may be accessible at} the LHC. It goes without saying that heavier masses for the exotic scalar fields can be achieved at the expense of the strict naturalness requirement we asked for.

\subsection{Ultralight pseudodilaton phenomenology \label{sec:dilatonpheno}}

The smoking gun of the current scenario at low energy is the light pseudodilaton, whose basic features are rather similar to those of the analogous state discussed in~\cite{Allison:2014zya,Allison:2014hna}. 

In particular, concerning the pseudodilaton couplings to the matter fields, they are driven by its mixing with the other neutral scalars. As one can see by inverting \eqs{tildeh0}{tildesigma0}, its projection onto the neutral components of the Higgs doublets amounts to $\sim v/V_\sigma$ which, in turn, yields couplings to the SM fermions of the order of $m_f/V_\sigma$. Although this, in principle, admits for a possible pseudodilaton detection in ``5th force'' experiments,  this interaction turns out to be too weak for the existing limits to provide nontrivial constraints. Since the relevant mediator mass $m_\sigma$ is of ${\cal O}(v^2/V_\sigma) \approx 30\times (10^9\,\GeV/V_\sigma)\, {\rm keV}$, the current bound reads $\alpha_5/\alpha_{EM}\lesssim 10^{-8} -10^{-16}$ \cite{Salumbides:2013dua}, for $V_\sigma$ in  the $10^{9} - 10^{12}$~GeV range, far above the size of the pseudodilaton interaction strength with ordinary matter. {The ultraweak size of the couplings of the complex singlet scalar field, which drives both the axion and the light pseudodilaton states, does not lead to visible collider signatures either, at variance with the case of less constrained scale invariant Higgs extensions~\cite{Farzinnia:2015uma}.}

Considering the role the pseudodilaton may play in the early Universe cosmology, the main concerns have to do with the energy stored in the coherent oscillations of the 
pseudodilaton field after inflation. To that end, two basic scenarios, characterized basically by the hierarchy between the reheat temperature and the PQ breaking scale, emerge~\cite{Allison:2014zya,Allison:2014hna}. It turns out, however, that in either of these cases there is a natural way to dissipate the excess energy either by the pseudodilaton interactions with the SM thermal bath or by taming its very production in the first place.  The former, in the current setup, may work even better than estimated in the large $\tan\beta$ scenario of~\cite{Allison:2014zya} since the small $\tan\beta$ constraint implies larger up-type Yukawa couplings and, in turn, more efficient dissipation. There is no natural domain for the relevant parameters in which the pseudodilaton may contribute a significant fraction of $\Omega_{DM}$ (unlike the axion): either its production is negligible or it is unstable on cosmological scales. 

Given all that, the main difference among our setup and that in~\cite{Allison:2014zya}, is the direct coupling of the pseudodilaton to the light neutrinos that follows from its mixing with the triplet scalar in the Type-II seesaw. 
On the other hand, the absolute size of the corresponding effective coupling, $g_{\sigma\nu\nu}\propto m_\nu/V_\sigma$, 
is utterly small and, at the present time, it yields no observable effect.

\subsection{Vacuum stability}

An added value of scalar extensions of the SM is their potential to improve on the stability of the electroweak vacuum. 
This issue has been discussed at length in the literature 
(see for instance Refs.~\cite{Lebedev:2012zw,EliasMiro:2012ay}), and we just briefly recall the argument here.
The key effect is the {positive} contribution of the new scalars (through, e.g., the Higgs portal couplings) 
to the {beta-function of the Higgs quartic coupling $\lambda_H$.  
As such, }
they tend to stabilize the Higgs potential if they enter the running below the instability 
scale.\footnote{The instability scale of the SM effective potential is a gauge dependent 
quantity \cite{DiLuzio:2014bua}. A gauge invariant criterium to include the effects of new 
physics can be devised~\cite{Andreassen:2014gha}.}

Nevertheless, the presence of multiple scalar field directions {may} potentially reintroduce 
the issue of instability already at tree level, especially in those cases where operators 
featuring an odd power of the same field are present (e.g.~those associated with the couplings 
$\lambda_5$ and $\lambda_6$ in \eq{TII-scalarpot}). 
On the other hand, {even at the tree-level potential level}, a fully analytical determination of the 
necessary and sufficient conditions {for the boundeness of \eq{TII-scalarpot} from below} 
turns out to be a formidable problem. 

\begin{figure}[t]
\includegraphics[width=0.5\textwidth]{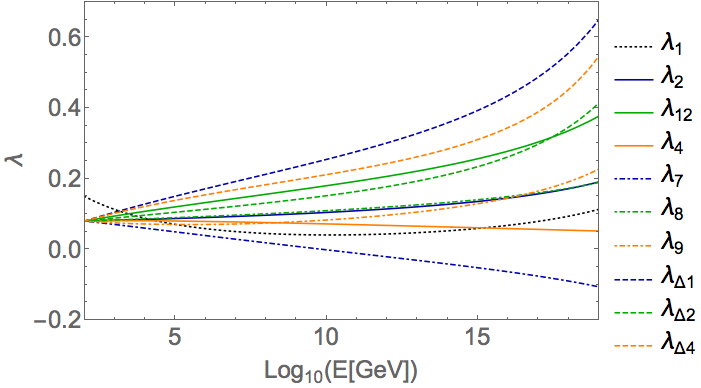} 
\caption{\label{runninglambdaL} One-loop running of the $\mathcal{O}(10^{-1})$ couplings for the weak-scale benchmark values specified in \Table{benchmarkL}.}
\end{figure}

In \fig{runninglambdaL} we display the one-loop running of the {``large'' (i.e., $\mathcal{O}(10^{-1})$) couplings for the weak-scale 
benchmark values} given in \Table{benchmarkL}. 

{Regardless of} the fact that {$\lambda_{4}$ and $\lambda_{7}$ both fall negative at large energies}, it can be shown that the relevant boundedness conditions are always satisfied. 
To this end, in \app{stability} we provide a compact parametrization of the scalar potential manifold  
based on the ``invariants' method'' (see e.g.~Ref.~\cite{Arhrib:2011uy}),  
which allows us to probe the {critical} scalar field directions by fully exploiting the symmetries of the system. 
In particular, the sixteen real field variables of the potential in \eq{TII-scalarpot} 
are traded for ten real parameters (three angles and six invariants defined over a compact domain 
plus one radial coordinate). 
In this way, given a set of benchmark values for the scalar potential couplings 
it is possible to perform a fast numerical check of the vacuum stability 
by randomly scanning over the scalar potential variables. 

\begin{figure}[t]
\includegraphics[width=0.49\textwidth]{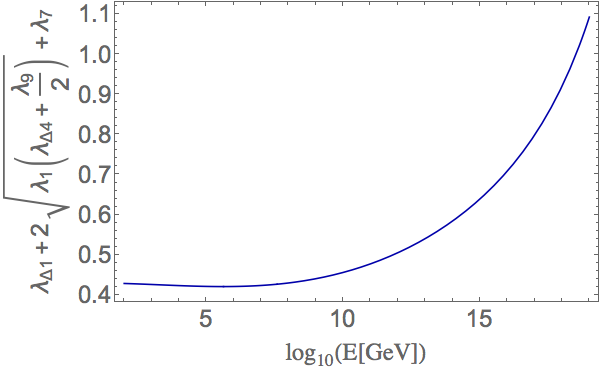} 
\caption{\label{lambdad17d42} One-loop running of the boundedness condition in \eq{runBCTII}, involving $\lambda_7$.}   
\end{figure}

Moreover, by selecting specific field directions (angles), 
{certain sufficient conditions for the vacuum} stability can be explicitly worked out (cf.~\app{stability}). 
This provides an analytical understanding of the reason why, {e.g.,
the fact that  $\lambda_7$ runs sharply negative, as shown in \fig{runninglambdaL},} 
is not necessarily a problem.
As a matter of fact, the relevant boundedness condition 
(in the Type-II seesaw limit -- cf.~\app{vacstaTII}) yields 
the relation (most restrictive for $\lambda_9>0$) 
\beq
\label{runBCTII}
 \lambda_{\Delta 1} + \lambda_7 + 2 \sqrt{\lambda_1 \left( \lambda_{\Delta 4} + \tfrac{1}{2} \lambda_9 \right)} > 0 \, ,
\eeq
which, as shown in \fig{lambdad17d42}, is always satisfied for the benchmark point in \Table{benchmarkL}. 
We have verified by analytic and numerical means that for the chosen set of benchmark values this conclusion holds for all field directions.

\section{Conclusions}
\label{conclusions}

We have discussed a minimal extension of the SM that addresses in a structural and natural way a number of its observational and theoretical open issues.
The origin of a Type-II seesaw neutrino mass is strictly related to the presence of an invisible axion \`a la DFSZ~\cite{Bertolini:2014aia}. In spite of the large PQ scale, a technically stable ultraweak setup of the singlet scalar couplings makes the model perturbatively stable and natural, with a scalar spectrum potentially {within} the reach of the LHC. The presence of the axion field {provides a natural solution} to both the dark matter issue and the strong CP problem. At the same time, the required extension of the scalar sector improves on the stability of the SM vacuum. We do not require additional fermions 
(recent proposals that include right-handed neutrinos in connection with an invisible axion 
were presented in~\cite{Dias:2006th,Dias:2014osa,Salvio:2015cja,Carvajal:2015dxa,Clarke:2015bea,Ahn:2015pia}, while an extensive list of studies on the subject is found in~\cite{Bertolini:2014aia}).

The embedding of the Type-II seesaw model, discussed in~\cite{Bertolini:2014aia}, in a classically scale invariant setup explicitly broken by perturbative quantum effects, allows us to further constrain the model and describe the scalar spectrum in terms of {just few} parameters. A setup consistent with {all the LHC constraints is naturally} obtained.
In addition to the invisible axion a very light pseudodilaton characterizes the present setup. Both fields exhibit tiny couplings to neutrinos, 
that, {however,  do not affect} present-day laboratory, astrophysical and cosmological {searches}. 

While not excluding additional ultraviolet physics, the model represents a minimal renormalizable extension of the SM that aims at answering, from a particle physics  perspective, some of the yet open issues of the standard electroweak theory. With this ambition, we are currently investigating whether
cold baryogenesis with axion dynamics at play~\cite{Servant:2014bla}, 
may find as well a structural embedding in such a context.

\section*{Acknowledgments}

\noindent

We acknowledge useful correspondence with F. Lyonnet during the process of validation of the code PyR@TE, that served as a check of our calculation of the scalar beta coefficients. S.B.~acknowledges partial support by the Italian MIUR Grant
No.~2010YJ2NYW001 and by the EU Marie Curie ITN UNILHC Grant No.~PITN-GA-2009-237920. 
The work of L.D.L.~is supported by the Marie Curie CIG program, project number PCIG13-GA-2013-618439. 
L.D.L.~is grateful to SISSA for hospitality and partial support during the development of this project. 
The work of H.K.~is supported by the Grant Agency of the Czech Technical University in Prague, Grant No.~SGS13/217/OHK4/3T/14. 
H.K. would like to thank SISSA for the hospitality during the early stages of the project.
The work of M.M. is supported by the Marie-Curie Career Integration Grant within the 7th European Community Framework Programme
FP7-PEOPLE-2011-CIG, Contract No. PCIG10-GA-2011-303565 and by the Foundation for support of science and research ``Neuron''.

\appendix

     \section{Scalar Mass spectrum}
     \label{scalarspectrum}

{The scalar spectrum} is readily obtained from the quadratic part of the shifted potential. 
The minimization of the CW potential in \eq{Veff}  {provides,} together with \eq{lambda3}  for $\lambda_3$, three additional relations among the quartic scalar couplings. 
We choose to write the three ultraweak couplings $\lambda_{i3}$ ($i=2,3,\Delta$) in terms of the remaining ones as 
\begin{widetext}
     \bea
     	\lambda_{13} &=& -\frac {(\lambda_{\Delta  1}+ c_{6}\cot\beta){v_{\Delta}}^2 +2\,\lambda_{{1}}{v_{{u}}}^{2} -c_{{5}}v^2\cot\beta
     		+\lambda_{{12}}{v_{{d}}}^{2}}{V_{\sigma}^2}
\label{lam13} \, ,
\\
     \lambda_{23} &=&  -{\frac {(\lambda_{{8}}+\lambda_{
     						{\Delta 2}}+ c_{6}\tan\beta){{ v_{\Delta}}}^{2}+2\,\lambda_{{
     						2}}{v_{{d}}}^{2}-c_{{5}}{v}^{2}\tan\beta+\lambda_{{12}}{v_{{u}}}^{2}}{{V_{{\sigma}}}^{2}}} 
\label{lam23} \, ,	
\\
     \lambda_{\Delta 3} &=& -{\frac {2\,(\lambda_{{9}}+\lambda_{{{\Delta  4}}}){{ v_{\Delta}}}^{2}+(\lambda_{{8}}+
     					\lambda_{{{\Delta  2}}})
     					{v_{{d}}}^{2}+\lambda_{{{\Delta  1}}}{v_{{u}}}^{2}+c_{{6}}v_{{d}}v_{{u}}}{V_\sigma^2}}  \, 
\label{lamD3} \, ,		
     \eea
  \end{widetext}     
where,  according to \eq{cnumbers}, we conveniently used the rescaled couplings $c_5$ and $c_6$.
Since the adopted ultraweak limit for the singlet interactions leads to a scalar mass spectrum clustered below the TeV scale, the minimization of {the} higher-order effects in the CW one-loop potential implies that the relevant scale for the stationarity equations is the weak scale. {From the known constraints on the scalar spectrum} this allows us to readily derive  the corresponding set of low-scale values of the scalar couplings {underpinning the vacuum stability analysis}. 
 
\subsection{Neutral scalars}     
     
With the help of \eqs{lam13}{lamD3} the mass matrix entries for the {neutral scalars} in the $\{  h_u^0,h_d^0 ,\sigma^0 , \delta^0   \}$ basis can {be} written as      
     \begin{align}
    M_S^2[1,1] &={\frac {4\,\lambda_{{1}}{v_{{u}}}^{3}+\lambda_{{5}}v_{{d}}{V_{{\sigma}}}^{2
   		}-\lambda_6 v_\Delta v_d V_\sigma}{v_{{u}}}} \, ,	\\
   	 M_S^2[1,2] &=M_S^2[2,1]=\lambda_6 v_\Delta V_\sigma -\lambda_{{5}}{V_{{\sigma}}}^{2}+2\,\lambda_{{12}}v_{{d}}v_{{u}} \, ,
     \end{align}	
    	
     \begin{align}
    M_S^2[1,3]&=M_S^2[3,1]= - \frac{1}{V_{\sigma}}(2\,{{ v_{\Delta}}}^{2}\lambda_{{{  \Delta 1}}}v_u  \nn  \\
               & +{ v_{\Delta}}\,
   		\lambda_{{6}}V_{{\sigma}}v_d 
   		+4\,\lambda_{{1}} {v_{{u}}^{3}}+2\,\lambda_{
   			{12}} {v_{{d}}}^{2}v_u ) \, , 
 \\
   		 M_S^2[1,4] &=M_S^2[4,1]=2\,\lambda_{{{  \Delta 1}}}{ v_{\Delta}}\,v_{{u}}+\lambda_{{6}}v_{{d}}V_{{\sigma}} \, ,		
     \end{align}	
     \begin{align}
   	 M_S^2[2,2] &={\frac {4\,\lambda_{{2}}{v_{{d}}}^{3}+\lambda_{{5}}{V_{{\sigma}}}^{2}v_{{u}
   			} -\lambda_6 v_\Delta v_u V_\sigma}{v_{{d}}}} \, , \\
   			 M_S^2[2,3] &=M_S^2[3,2]=-\frac{1}{V_{\sigma}}(2\,{{ v_{\Delta}}}^{2} (\lambda_{{8}} +
   					\lambda_{{{  \Delta 2}}})v_{{d}}\nonumber \\
   					&+{ v_{\Delta}}\,\lambda_{{6}}V_{{\sigma}}v_{{u}}+4\,
   					\lambda_{{2}}{v_{{d}}}^{3}+2\,\lambda_{{12}}v_{{d}}{v_{{u}}}^{2})  \, ,
    \\
 M_S^2[2,4] &=M_S^2[4,2]= 2\,(\lambda_{{8}}+\lambda_{{{  \Delta 2}}}) v_{{d}}{ 
	v_{\Delta}}+\lambda_{{6}}V_{{\sigma}}v_{{u}} \, ,
     \end{align}		
     \begin{align}
 M_S^2[3,3] &= 2\, \beta_{\lambda_3}V_{\sigma}^2 +2\lambda_3^\prime v^2 -\lambda_6\frac{v_\Delta v_u v_d}{V_\sigma} \, ,
	\\
 M_S^2[3,4] &=M_S^2[4,3]=2\,\lambda_{{{  \Delta 3}}}{ v_{\Delta}}\,V_{{\sigma}}+\lambda_{{6}}v_{{d}}v_{{u}} \, ,     \\
 M_S^2[4,4] &= 4\,(\lambda_{{9}} + \lambda_{{{  \Delta 4}}}){v_{{\Delta}}}^{2}-\,\lambda_{{6}}\frac{v_{{d}}V_{{\sigma}}v_{{u}}}{{  v_{\Delta}}} \, ,
     \end{align}
where $\beta_{\lambda_3}$ is the one-loop beta function of the $\lambda_3$ coupling given in \app{1looprge}      
and $\lambda_3^\prime v^2$ is defined in \eq{veps}.
     
 {By neglecting} the mixings with the triplet ($v_\Delta \ll v \ll
 V_\sigma$) and by taking into account the ultraweak limit in \eq{UWlimit}, the {neutral scalar} mass matrix reduces to:
   \begin{widetext}
   	\begin{eqnarray}
   	M_S^{2}=	\left( \begin{array}{ccccc}
   	4\lambda_1v_u^2+c_5 v^2\cot\beta
   	& -c_5 v^2+2\,\lambda_{{12}}v_{{d}}v_{{u}}
   	&-\frac{1}{V_{\sigma}}(4\,\lambda_{{1}}{v_{{u}}}^{3}+2\,\lambda_{
   		{12}}v_{{d}}^{2}v_u)  & 0 \\
   	-c_5 v^2+2\,\lambda_{{12}}v_{{d}}v_{{u}} &  4\lambda_2v_d^2+c_5v^2\tan\beta & -\frac{1}{V_{\sigma}}(4\,
   	\lambda_{{2}}{v_{{d}}}^{3}+2\,\lambda_{{12}}v_{{d}}{v_{{u}}}^{2} )& 0  \\
   -\frac{1}{V_{\sigma}}( 4\,\lambda_{{1}}{v_{{u}}}^{3}+2\,\lambda_{
   {12}}v_{{d}}^{2}v_u) &  -\frac{1}{V_{\sigma}}(4\,
   	\lambda_{{2}}{v_{{d}}}^{3}+2\,\lambda_{{12}}v_{{d}}{v_{{u}}}^{2} )&2\beta_{\lambda_3}V_{\sigma}^2+2\lambda_3^\prime v^2 & 0 \\
   	0 & 0 &0 & -c_6v_uv_d 
   	\end{array} \right)   .
	\label{mneutral}
   	\end{eqnarray} 
   \end{widetext}
   
\eq{mneutral} can be conveniently written as
 \begin{eqnarray}
 M_S^{2}=	\left( \begin{array}{ccc}
 M^2_{\text{2H}} & V &  0  \\ 
 V^T &2 \beta_{\lambda_3}V_{\sigma}^2+2\lambda_3^\prime v^2 &  0 \\
  0 & 0 & -c_6v_uv_d
 \end{array} \right)   ,
 \label{MS2}
 \end{eqnarray}
 where
 	\begin{eqnarray}
 M^2_{\text{2H}} \equiv	\left( \begin{array}{ccccc}
 	4\lambda_1v_u^2+c_5 v^2\cot\beta  & -c_5 v^2+2\,\lambda_{{12}}v_{{d}}v_{{u}}\\
	-c_5 v^2+2\,\lambda_{{12}}v_{{d}}v_{{u}} &  4\lambda_2v_d^2+c_5v^2\tan\beta  
 	\end{array} \right) \nonumber \\
 	\end{eqnarray}
and
 		\begin{eqnarray}
 		V  \equiv  -\frac{1}{V_{\sigma}}	\left( \begin{array}{ccccc}
 4\,\lambda_{{1}}{v_{{u}}}^{3}+2\,\lambda_{
 		{12}}v_{{d}}^{2}v_u
 		 \\
 		4\,
 		 \lambda_{{2}}{v_{{d}}}^{3}+2\,\lambda_{{12}}v_{{d}}{v_{{u}}}^{2} 
 		\end{array} \right) . \nonumber \\
 		\end{eqnarray}
 The mass of the (mainly) singlet eigenstate $\tilde{\sigma}$ is readily given at $O(v^4/V_\sigma^4)$  by 
 	\begin{equation}
 	 m^2_{\tilde{\sigma}}=  2\, \beta_{\lambda_3}V_{\sigma}^2+2\lambda_3^\prime v^2 - V^T ( M^2_{\text{2H}})^{-1} V  \, .
 	\end{equation}
{Using  \eq{veps2} a straightforward computation yields}
 	\begin{align}
  V^T ( M^2_{\text{2H}})^{-1} V =&\ \frac{4}{V_{\sigma}^2}(\lambda_1v_u^4+\lambda_2 v_d^4+\lambda_{12}v_u^2v_d^2) \nonumber \\
  =& \ 2 \lambda_3^\prime v^2 \, ,
 	\end{align}
 	from where we obtain the pseudodilaton mass
 	\begin{equation}
 	m^2_{\tilde{\sigma}}= 2\, \beta_{\lambda_3}V_{\sigma}^2 ,
 	\end{equation} 
{which, as expected, turns out to be} proportional to the explicit breaking of {the} scale
        invariance induced by the running of $\lambda_3$ in the stress-tensor trace anomaly. This result obviously holds for the complete mass matrix as well. 
	
The mass eigenvalues of the light $\tilde{h}$ and heavy $\tilde{H}$ neutral scalars are then given by
 	\begin{widetext}
 	\beq
 	2\, m_{\tilde{h},\tilde{H}}^2\approx\left[4\lambda_1v_u^2+4\lambda_2v_d^2+c_5 v^2(\tan\beta+\cot\beta) \right] \mp \sqrt{ \left[4\lambda_2v_d^2-4\lambda_1v_u^2 +c_5 v^2(\tan\beta-\cot\beta) \right]^2+4(2\lambda_{12}v_uv_d-c_5 v^2)^2}
	\label{hHtildemasses}  \, .
 	\eeq
 	\end{widetext}

Up to the {doublet-triplet mixing proportional, in the ultraweak
  setup,} to $v_\Delta/v$ ($< 1\%$) and to the mixings with the singlet field, suppressed by the PQ scale, the corresponding mass eigenstates are {written} as 
 	\begin{align}
 	& \tilde{h} \approx \cos \alpha\, h_u^0 -\sin\alpha\, h_d^0  \label{tildeh} \, ,\\ 
 	& \tilde{H} \approx \sin \alpha\, h_u^0 +\cos\alpha\, h_d^0   \label{tildeH} \, , \\
 	& \tilde{\sigma}_0 \approx \sigma_0  \label{tildesigma} \, , \\
 	& \tilde{\delta}_0 \approx \delta_0  \, ,
	\label{eigenstates4}
 	\end{align}
 	where $\alpha$ is the mixing angle between the neutral doublet fields, given by
     \begin{equation}
     	\tan 2 \alpha =  \frac{2 (2\lambda_{12}v_uv_d-c_5 v^2)}{4(\lambda_2v_d^2-\lambda_1v_u^2)+c_5(\tan\beta-\cot\beta)v^2}\, ,
	\label{alphamix}
     \end{equation}
    with $-\pi/2 <\alpha< \pi/2$. \eqs{tildeh}{alphamix} are consistent with the {two-doublet} mixing and eigenstate definitions of Ref.~\cite{Gunion:2002zf}. 

Were the triplet mixings negligible with respect to those {of the singlet sector} ($v_\Delta / v \ll v / V_\sigma$) the eigenstates would read
 	\begin{align}
 	& \tilde{h} \approx \cos \alpha(\, h_u^0  -\frac{v_u}{V_{\sigma} }  \sigma_0) - \sin\alpha( h_d^0  -\frac{v_d}{V_{\sigma} }\sigma_0) \label{tildeh0} \, ,\\ 
 	& \tilde{H} \approx \sin \alpha(\, h_u^0  -\frac{v_u}{V_{\sigma} }  \sigma_0) +\cos \alpha (\, h_d^0  -\frac{v_d}{V_{\sigma} }  \sigma_0) \label{tildeH0} \, , \\
 	& \tilde{\sigma}_0 \approx \sigma_0 + \frac{v_u}{V_{\sigma}}h_u^0+\frac{v_d}{V_{\sigma}}h_d^0 \label{tildesigma0} \, ,
 	\end{align}
where the result
 	\beq
 	( M^2_{\text{2H}})^{-1} V =-\frac{1}{V_{\sigma}}	\left( \begin{array}{ccccc}
 	v_u   \\ 
     	v_d
 	\end{array} \right).
 	\eeq
is used in the seesaw-like diagonalization of $M_S^2$ in \eq{MS2}.

     \subsection{Pseudoscalar fields}
     
 Analogously, for the pseudoscalar fields we obtain in the $(  \eta_u^0, \eta_d^0 ,\eta_\sigma^0 , \eta_\delta^0)$ basis

     \begin{align}
     & M_\text{PS}^2[1,1]=-{\frac {v_{{d}}{V_{{\sigma}}} \left( {  v_{\Delta}}\,\lambda_{{6}}-\lambda_{{5}} V_{{\sigma}} \right) }{v_{{u}}}} \, ,
     \\
     & M_\text{PS}^2[1,2]=M_\text{PS}^2[2,1]=\lambda_{{6}}{  v_{\Delta}}\,{V_{{\sigma}}}+\lambda_{{5}}V_{{\sigma}}^{2} \, ,\\
     & M_\text{PS}^2[1,3]=M_\text{PS}^2[3,1]=\lambda_{{6}}{  v_{\Delta}}\,v_{{d}}+2\,\lambda_{{5}}v_{{d}}{V_{{\sigma}}} \, , \\
     & M_\text{PS}^2[1,4]=M_\text{PS}^2[4,1]=-\lambda_{{6}}v_{{d}}{V_{{\sigma}}} \, , \\
     & M_\text{PS}^2[2,2]=-{\frac {V_{{\sigma}}v_{{u}} \left( { v_{\Delta}}\,\lambda_{{6}}-\lambda_{{5}}V_{{\sigma}} \right) }{v_{{d}}}} \, ,
    \\
     & M_\text{PS}^2[2,3]=M_\text{PS}^2[3,2]=-\lambda_{{6}}{  v_{\Delta}}\,v_{{u}}+2\,\lambda_{{5}}V_{{\sigma}}v_{{u}} \, ,
     \\
     & M_\text{PS}^2[2,4]=M_\text{PS}^2[4,2]=\lambda_{{6}}V_{{\sigma}}v_{{u}} \, ,
     \\
     & M_\text{PS}^2[3,3]= -\lambda_6\frac{v_{\Delta}v_uv_d}{V_{\sigma}}+4\lambda_5v_uv_d \, ,
     \\
     & M_\text{PS}^2[3,4]=M_\text{PS}^2[4,3]=\lambda_{{6}}v_{{d}}v_{{u}} \, ,\\ 
     & M_\text{PS}^2[4,4]=-{\frac {\lambda_{{6}}V_{{\sigma}}v_{{d}}v_{{u}}}{{  v_{\Delta}}}} \, .
     \end{align}
     After diagonalization and by neglecting  the $v_\Delta$ terms, we finally get     
     \begin{eqnarray}
     M_\text{PS}^{2\text{}}=	\left( \begin{array}{ccccc}
     0
     &0
     &0 & 0 \\
     0 & c_5 \frac{v^4}{v_u v_d} & 0 & 0 \\
     0 & 0&0 & 0 \\
     0 & 0 &0 &-c_6 v_uv_d 
     \end{array} \right) \, ,
     \end{eqnarray}
   where 
$c_{5,6}$ are {${\cal O}(1)$} numbers in the ultraweak limit. The two massless states correspond to the weak neutral GB and to the invisible axion {fields,} respectively. The latter acquires a tiny mass via QCD instantons.
     
     \subsection{Singly charged scalars}
 
    {Working in} the $\{  h_u^+,h_d^+ , \delta^+   \}$ basis we obtain
          
     \begin{align}
     & M_+^2[1,1]=\lambda_{{7}}{{  v_{{\Delta }}}}^{2}-{  v_{{\Delta }}}\,\lambda_{{6}}\frac{v_
     	{{d}}}{v_{{u}}}V_{{\sigma}}+\lambda_{{4}}{v_{{d}}}^{2}+\lambda_{{5}}\frac{v_{{d}}}{v_{{u}}}{V_{{\sigma}}}^{2} \, ,
     \\
     & M_+^2[1,2]=M_+^2[2,1]=\lambda_{{4}}v_{{d}}v_{{u}}+\lambda_{{5}}{V_{{\sigma}}}^{2} \, ,\\
     & M_+^2[1,3]=M_+^2[3,1]=\frac{1}{\sqrt{2}}(\,\lambda_{{7}}v_{{u}}{  v_{{\Delta }}}-\,\lambda_{{6}}v_{{d}}V_{{\sigma}}) \, ,\\ 
     & M_+^2[2,2]= -v_{{\Delta }}^{2}\lambda_{{8}}-{  v_{{\Delta }}}V_{{\sigma}}\,\lambda_{{6}}\frac{v_{{u}}}{v_{{d}}}+\lambda_{{4}}{v_{{u}}}^{2}+\lambda_{{5}}V_{{\sigma}}
     ^{2}\frac{v_{{u}}}{v_{{d}}} \, ,
     \\
     & M_+^2[2,3]=M_+^2[3,2]=\frac{1}{\sqrt{2}}(\,\lambda_{{8}}{  v_{{\Delta }}}\,v_{{d}}+\,\lambda_{{6}}V_{{\sigma}}v_{{u}}) \, , \\
     & M_+^2[3,3]=\frac{1}{2}(\lambda_{{7}}{v_{{u}}}^{2}-\lambda_{{8}}{v_{{d}}
     }^{2})-\,\lambda_{{6}}\frac{V_{{\sigma}}v_{{d}}v_{{u}}}{{  v_{{\Delta }}}} \, .
     \end{align}
     
     After diagonalization and by neglecting  $v_\Delta$ terms, we finally get
     
     	\beq
     	M_{+}^{2}  =	\left( \begin{array}{ccccc}
     	0     	&0     	&0  \\
     	0 &\lambda_4v^2+ c_5 \frac{v^4}{v_u v_d} & 0 \\
     	0 & 0  &\frac{1}{2}(\lambda_{{7}}{v_{{u}}}^{2}-\lambda_{{8}}{v_{{d}}
     	}^{2}) -\,c_{{6}}v_{{d}}v_{{u}}
     	\end{array} \right) \, ,
     	\eeq

\noindent where the massless state corresponds to the weak charged GB.

     \subsection{Doubly charged scalars}	
 
  The mass of the doubly charged component of the scalar triplet {reads}  
     \beq
     M^2_{++}=\lambda_{{7}}{v_{{u}}}^{2}-\lambda_{{8}}{v_{{d}}}^{2}
     -c_{{6}}v_{{d}}v_{{u}}\ .
     \eeq
Notice the $\tfrac{1}{2}(\lambda_{{7}}{v_{{u}}}^{2}-\lambda_{{8}}{v_{{d}}}^{2})$ mass isosplitting among the components of the scalar triplet.

\section{Scalar one-loop beta functions}
\label{1looprge}

We consider the field content given in Table \ref{fctypeII} with
the scalar potential in \eq{TII-scalarpot}
and the Yukawa Lagrangian in \eq{YukawaTII}. 
For the purpose of the present discussion we neglect the $Y_d$, $Y_e$, and $Y_\Delta$ couplings and take
\beq
Y_u \approx \left(\begin{array}{ccc}
0&0&0\\
0&0&0\\
0&0&y_t
\end{array}\right).
\eeq
{As usual, we denote by  $g_3$, $g_2$ and $g_1$ the gauge couplings
  associated to the three factor of the $SU(3)_C\otimes SU(2)_L\otimes
  U(1)_Y$ gauge group, respectively.} 
In {this} approximation the one-loop running of the top Yukawa coupling coincides with the SM result, namely
\beq
\frac{\mathrm{d}y_t}{\mathrm{d}\ln\mu} =
 \frac{y_t}{16 \pi^2} (9/2 y_t^2 - 8 g_3^2 - 9/4 g_2^2 - 17/12 g_1^2) \, .
\eeq
On the other hand, the gauge coupling running is modified by the presence of the additional scalars:
\beq
\frac{\mathrm{d}\alpha_i^{-1}}{\mathrm{d}\ln\mu}=-\frac{a_i}{2\pi} \ ,
\eeq
with $a=\{8, -\frac{7}{3}, -7\}$ and $\alpha_i \equiv g^2_i / (4 \pi)$.

{Parametrizing the $\beta$ functions of the scalar couplings as
\beq
\frac{\mathrm{d}\lambda}{\mathrm{d}\ln{\mu}} = \frac{b_\lambda}{16\pi^2} \equiv \beta_\lambda\ ,
\eeq
we} obtain the following one-loop {results~\cite{Machacek:1984zw}:}
\bwt
\bea
b_{\lambda_1}  &=& 
24 \lambda_1^2 + 2 \lambda_{12}^2 + 2 \lambda_{12} \lambda_{4} + \lambda_{4}^2 +
  \lambda_{13}^2 + 3 \lambda_{\Delta 1}^2 + 3 \lambda_{\Delta 1} \lambda_{7} + 5/4 \lambda_{7}^2
  - 6 y_t^4  + 12 y_t^2 \lambda_1 \nn\\
&&  + 3/8 g_1^4 + 3/4 g_1^2 g_2^2 + 9/8 g_2^4  - 3 \lambda_1 (g_1^2 + 3 g_2^2),\\[1ex]
b_{\lambda_2}  &=& 
24 \lambda_2^2 + 2 \lambda_{12}^2 + 2 \lambda_{12} \lambda_{4} + \lambda_{4}^2 +
  \lambda_{23}^2 + 3 \lambda_{\Delta 2}^2 + 3 \lambda_{\Delta 2} \lambda_{8} + 5/4 \lambda_{8}^2\nn\\
 && + 3/8 g_1^4 + 3/4 g_1^2 g_2^2 + 9/8 g_2^4
  - 3 \lambda_2 (g_1^2 + 3 g_2^2),\\[1ex]
b_{\lambda_{12}}  &=& 
12 \lambda_1 \lambda_{12} + 4 \lambda_1 \lambda_{4} + 12 \lambda_2 \lambda_{12} +
  4 \lambda_2 \lambda_{4} + 2 \lambda_{13} \lambda_{23} + 6 \lambda_{\Delta 1} \lambda_{\Delta 2} +
  3 \lambda_{\Delta 1} \lambda_{8} + 3 \lambda_{\Delta 2} \lambda_{7} \nn\\
  &&+ 1/2 \lambda_{7} \lambda_{8} +
  4 \lambda_{12}^2 + 2 \lambda_{4}^2 + 4 \lambda_{5}^2 + 2 \lambda_{6}^2
  + 6 y_t^2 \lambda_{12}
  + 3/4 g_1^4 + 3/2 g_1^2 g_2^2 + 9/4 g_2^4
  - 3 \lambda_{12} (g_1^2 + 3 g_2^2),\\[1ex]
b_{\lambda_4}  &=& 
4 \lambda_{4} (\lambda_1 + \lambda_2 + 2 \lambda_{12} + \lambda_{4}) +
  2 \lambda_{7} \lambda_{8} - 4 \lambda_{5}^2 - \lambda_{6}^2
  + 6 y_t^2 \lambda_{4}
  - 3 g_1^2 g_2^2
  - 3 \lambda_{4} (g_1^2 + 3 g_2^2),\\[1ex]
b_{\lambda_3}  &=& 
20 \lambda_{3}^2 + 2 \lambda_{13}^2 + 2 \lambda_{23}^2 + 4 \lambda_{5}^2 +
  3 \lambda_{\Delta 3}^2,\\[1ex]
b_{\lambda_{13}}  &=& 
8 \lambda_{3} \lambda_{13} + 12 \lambda_1 \lambda_{13} + 4 \lambda_{23} \lambda_{12} +
  2 \lambda_{23} \lambda_{4} + 6 \lambda_{\Delta 3} \lambda_{\Delta 1} + 3 \lambda_{\Delta 3} \lambda_{7} +
  4 \lambda_{13}^2 + 8 \lambda_{5}^2 + 3 \lambda_{6}^2\nn\\
  &&+ 6 y_t^2 \lambda_{13}
  - 3/2 \lambda_{13} (g_1^2 + 3 g_2^2),\\[1ex]
b_{\lambda_{23}}  &=& 
8 \lambda_{3} \lambda_{23} + 12 \lambda_2 \lambda_{23} + 4 \lambda_{13} \lambda_{12} +
  2 \lambda_{13} \lambda_{4} + 6 \lambda_{\Delta 3} \lambda_{\Delta 2} + 3 \lambda_{\Delta 3} \lambda_{8} +
  4 \lambda_{23}^2 + 8 \lambda_{5}^2 + 3 \lambda_{6}^2\nn\\
  &&- 3/2 \lambda_{23} (g_1^2 + 3 g_2^2),\\[1ex]
%
b_{\lambda_{\Delta 1}}  &=& 
12 \lambda_1 \lambda_{\Delta 1} + 4 \lambda_1 \lambda_{7} + 4 \lambda_{12} \lambda_{\Delta 2} +
  2 \lambda_{4} \lambda_{\Delta 2} + 2 \lambda_{12} \lambda_{8} + 2 \lambda_{13} \lambda_{\Delta 3} +
  16 \lambda_{\Delta 1} \lambda_{\Delta 4} \nn\\
  &&+12 \lambda_{\Delta 1} \lambda_{9} + 6 \lambda_{7} \lambda_{\Delta 4} +
  2 \lambda_{7} \lambda_{9} + 4 \lambda_{\Delta 1}^2 + \lambda_{7}^2  + 2 \lambda_{6}^2
  + 6 y_t^2 \lambda_{\Delta 1}\nn\\
  &&+ 3 g_1^4  + 6 g_1^2 g_2^2 + 6 g_2^4
  - 3/2 \lambda_{\Delta 1} (5 g_1^2 + 11 g_2^2),\\[1ex]
b_{\lambda_{\Delta 2}}  &=& 
12 \lambda_2 \lambda_{\Delta 2} + 4 \lambda_2 \lambda_{8} + 4 \lambda_{12} \lambda_{\Delta 1} +
  2 \lambda_{4} \lambda_{\Delta 1} + 2 \lambda_{12} \lambda_{7} + 2 \lambda_{23} \lambda_{\Delta 3} +
  16 \lambda_{\Delta 2} \lambda_{\Delta 4} \nn\\
  &&+12 \lambda_{\Delta 2} \lambda_{9} + 6 \lambda_{8} \lambda_{\Delta 4} +
  2 \lambda_{8} \lambda_{9} + 4 \lambda_{\Delta 2}^2 + \lambda_{8}^2
  + 3 g_1^4  - 6 g_1^2 g_2^2 + 6 g_2^4
  - 3/2 \lambda_{\Delta 2} (5 g_1^2 + 11 g_2^2),\\[1ex]
b_{\lambda_{\Delta 3}}  &=& 
8 \lambda_{3} \lambda_{\Delta 3} + \lambda_{\Delta 3} (16 \lambda_{\Delta 4} + 12 \lambda_{9}) +
  2 \lambda_{13} (2 \lambda_{\Delta 1} + \lambda_{7}) +
  2 \lambda_{23} (2 \lambda_{\Delta 2} + \lambda_{8}) \nn\\
&&+ 4 \lambda_{\Delta 3}^2 + 2 \lambda_{6}^2
  - 3 \lambda_{\Delta 3} (2 g_1^2 + 4 g_2^2),\\[1ex]
b_{\lambda_{\Delta 4}}  &=& 
28 \lambda_{\Delta 4}^2 + 24 \lambda_{\Delta 4} \lambda_{9} + 6 \lambda_{9}^2 +
  2 \lambda_{\Delta 1} (\lambda_{\Delta 1} + \lambda_{7}) + 2 \lambda_{\Delta 2} (\lambda_{\Delta 2} + \lambda_{8}) +
  \lambda_{\Delta 3}^2\nn\\
  &&+ 6 g_1^4  - 12 g_1^2 g_2^2 + 15 g_2^4 - 3 \lambda_{\Delta 4} (4 g_1^2 + 8 g_2^2)),\\[1ex]
%
b_{\lambda_{5}}  &=& 
\lambda_{5} (4 \lambda_{3} + 2\lambda_{12} + 4\lambda_{13} + 4\lambda_{23} - 2\lambda_{4})
  + 3 y_t^2 \lambda_{5}
  - 3/2 \lambda_{5} (g_1^2 + 3 g_2^2),\\[1ex]
b_{\lambda_{6}}  &=& 
2 \lambda_{6} (\lambda_{13} + \lambda_{23} + \lambda_{\Delta 2} + \lambda_{\Delta 1}) +
  \lambda_{6} (3 \lambda_{8} - \lambda_{7}) + 2 \lambda_{6} \lambda_{\Delta 3} +
  2 \lambda_{6} \lambda_{12}
  + 3 y_t^2 \lambda_{6}
  - 3/2 \lambda_{6} (3g_1^2 + 7g_2^2),\\[1ex]
b_{\lambda_{7}}  &=& 
4 \lambda_1 \lambda_{7} + 4 \lambda_{\Delta 4} \lambda_{7} + 8 \lambda_{7} \lambda_{9} +
  2 \lambda_{4} \lambda_{8} + 8 \lambda_{\Delta 1} \lambda_{7} + 4 \lambda_{7}^2 - 2 \lambda_{6}^2
  + 6 y_t^2 \lambda_{7} \nn\\
  &&- 12 g_1^2 g_2^2
  - 3/2 \lambda_{7} (5 g_1^2 + 11 g_2^2),\\[1ex]
b_{\lambda_{8}}  &=& 
4 \lambda_2 \lambda_{8} + 4 \lambda_{\Delta 4} \lambda_{8} + 8 \lambda_{8} \lambda_{9} +
  2 \lambda_{4} \lambda_{7} + 8 \lambda_{\Delta 2} \lambda_{8} + 4 \lambda_{8}^2 + 2 \lambda_{6}^2
  + 12 g_1^2 g_2^2
  - 3/2 \lambda_{8} (5 g_1^2 + 11 g_2^2),\\[1ex]
b_{\lambda_{9}}  &=& 
24 \lambda_{\Delta 4} \lambda_{9} + 18 \lambda_{9}^2 + \lambda_{7}^2 + \lambda_{8}^2
  + 24 g_1^2 g_2^2 - 6 g_2^4
  - 3 \lambda_{9} (4 g_1^2 + 8 g_2^2) \, .
\eea
\ewt

Notice that $\lambda_5$ and $\lambda_6$ are individually multiplicatively renormalized and, together with the $\lambda_{i3}$ couplings, renormalize multiplicatively as a subset. These results are expected from symmetry arguments and hold at any order in perturbation theory, 
thus making the ultraweak limit technically natural. {As expected}, the quartic singlet coupling $\lambda_3$ scales quadratically {with the ultraweak couplings which  makes it natural to assume the dominance of loop corrections in the CW} potential in the singlet direction.

\section{Invariants' method for the vacuum stability analysis}
\label{stability}

In order to study the vacuum stability of the potential in \eq{TII-scalarpot} 
we employ the following polar representation
\begin{align}
|H_u| &= r \sin\chi\sin\theta\cos\phi  \, , \\
|H_d| &= r \sin\chi\sin\theta\sin\phi  \, , \\
|\sigma| &= r \sin\chi\cos\theta \, , \\
|\Delta| &= r \cos\chi \, , 
\end{align}
with $r \in [0,\infty)$, $\chi \in [0,\frac{\pi}{2}]$, $\theta \in [0,\frac{\pi}{2}]$, $\phi \in [0,\frac{\pi}{2}]$
and define the (normalized) invariants
\begin{align}
\zeta_4 &=\frac{|H^\dag_u H_d|^2}{|H_u|^2 |H_d|^2} \, , \\
\zeta_5 &=\frac{\text{Re}( \sigma^2 \tilde{H}^\dag_u H_d )}{|\sigma|^2 |H_u| |H_d|} \, ,\\
\zeta_6 &=\frac{\text{Re}( \sigma {H}^\dag_u \Delta^\dag H_d )}{|\sigma| |H_u| |\Delta| |H_d|} \, ,  
\end{align}
\begin{align}
\zeta_7 &=\frac{{H}^\dag_u \Delta \Delta^\dag H_u}{|H_u|^2 |\Delta|^2} \, , 
\end{align} 
\begin{align}
\zeta_8 &=\frac{{H}^\dag_d \Delta \Delta^\dag H_d}{|H_d|^2 |\Delta|^2} \, ,  
\end{align} 
\begin{align}
\zeta_9 &=\frac{\Tr (\Delta^\dag \Delta \Delta^\dag \Delta)}{|\Delta|^4} \, , 
\end{align} 
\vfill
where
$\zeta_4 \in [0,1]$, $\zeta_5 \in [-1,1]$, $\zeta_6 \in [-1,1]$, $\zeta_7 \in [0,1]$, $\zeta_8 \in [0,1]$, 
$\zeta_9 \in [\tfrac{1}{2},1]$. 

The {domains of the invariants follow} 
from the Cauchy-Schwarz inequality.\footnote{In order to determine the domain of 
$\zeta_9$ it is useful to reexpress it as 
$\zeta_9 = 1 - \frac{1}{2} \frac{({\bf \Delta}^\dag \cdot {\bf \Delta}^\dag)({\bf \Delta} \cdot {\bf \Delta})}{|{\bf \Delta}|^4}$, 
where we used the definition in \eq{expdelta} and
$\Tr (\tau_i\tau_j\tau_k\tau_l) = 2 (\delta_{ij}\delta_{kl}+\delta_{il}\delta_{jk}-\delta_{ik}\delta_{jl})$.} 
Note that since the coefficients $\zeta_{4,\ldots,9}$ are correlated, their allowed range of variation 
is, {as a whole,} reduced~\cite{Bonilla:2015eha}. 
However, since we are mainly interested in the sufficient conditions for stability we will let $\zeta_{4,\ldots,9}$ 
vary independently. 

By plugging the above parametrization {into} the scalar potential of \eq{TII-scalarpot} we get
\begin{widetext}
\begin{align}
V_0 &= r^4 
\Big\{ 
\left( \lambda_{\Delta 4} + \zeta_9 \lambda_9 \right) \cos^4\chi  + \left[ \lambda_{\Delta 3} \cos^2\theta 
+ \left( (\lambda_{\Delta 1} + \zeta_7 \lambda_7 ) \cos^2\phi + (\lambda_{\Delta 2} + \zeta_8 \lambda_8 ) \sin^2\phi \right) \sin^2\theta \right]  \cos^2\chi \sin^2\chi \nonumber \\
&+ 2 \zeta_6 \lambda_6   \cos \phi \sin \phi \cos\theta \sin^2\theta \cos\chi \sin^3\chi + \left[ \lambda_3 \cos^4\theta 
+ \left( \lambda_{13} \cos^2\phi + \lambda_{23} \sin^2\phi + 2 \zeta_5 \lambda_5 \cos\phi \sin\phi \right) \cos^2\theta \sin^2\theta \right. \nonumber \\
& \left. + \left( \lambda_{1} \cos^4\phi + \lambda_{2} \sin^4\phi + (\lambda_{12} + \zeta_4 \lambda_4) \cos^2\phi \sin^2\phi \right) \sin^4\theta \right] \sin^4\chi
\Big\} \, . 
\end{align}
\end{widetext}
The (tree-level) stability problem, $V_0 > 0$, is hence reduced to an inequality in terms of nine real variables 
spanning over a compact domain. Given a benchmark set of the scalar couplings 
({e.g.,~the} one in \Table{benchmarkL}), 
we can check that the potential is bounded from below 
by means of a numerical scan over the potential parameters, which consist of three angles and six invariants. 

In the following, we consider few specific field directions where  the (sufficient) stability 
conditions in some limiting cases can be determined analytically.

\subsection{DFSZ limit: $(\chi = \frac{\pi}{2})$}

In this case the field $\Delta$ is decoupled from the DFSZ model and the tree-level potential is given by 
\begin{widetext}
\begin{align}
\label{scalarpotpar}
V_0 = 
r^4 \cos^4\theta \left\{ \lambda_3 \right.
&\left. + \left( \lambda_{13} \cos^2\phi +  \lambda_{23} \sin^2\phi + 2 \zeta_5 \lambda_5 \cos\phi \sin\phi \right) 
\tan^2\theta \right. 
\nonumber \\
& \left. 
+ \left[ \lambda_1 \cos^4\phi + \lambda_2 \sin^4\phi + \left( \lambda_{12} + \zeta_4 \lambda_4 \right) \cos^2\phi \sin^2\phi \right] \tan^4\theta
\right\}
\, ,  
\end{align}
\end{widetext}
The expression in the curly bracket is a biquadratic function of {$\tan\theta \in [0,\infty)$; this readily gives the following} positivity
constraints:\footnote{Given the polynomial
$V(\chi) = a + b x^2 + c x^4$, 
with $x \in [0,\infty)$, the condition $V(x) > 0$ yields $a>0$, $c>0$ and $b + 2 \sqrt{ac} > 0$.
}
\begin{widetext}
\begin{align}
i) \ & \lambda_3 > 0 \, , \\
ii) \ & \lambda_1 \cos^4\phi + \lambda_2 \sin^4\phi + \left( \lambda_{12} + \zeta_4 \lambda_4 \right) \cos^2\phi \sin^2\phi > 0 \, , \\
iii) \ & \lambda_{13} \cos^2\phi +  \lambda_{23} \sin^2\phi + 2 \zeta_5 \lambda_5 \cos\phi \sin\phi 
 + 2 \sqrt{\lambda_3 \left( \lambda_1 \cos^4\phi + \lambda_2 \sin^4\phi + \left( \lambda_{12} + \zeta_4 \lambda_4 \right) \cos^2\phi \sin^2\phi 
\right)} > 0 \, . 
\end{align}
\end{widetext}

Let us consider {the} conditions $ii)$ and $iii)$. By factorizing $\cos^4\phi$ out of $ii)$ we get again a biquadratic form 
in $\tan\phi \in [0,\infty)$
\begin{equation}
\lambda_1 + \left( \lambda_{12} + \zeta_4 \lambda_4 \right) \tan^2\phi + \lambda_2 \tan^4\phi > 0 \, , 
\end{equation}
which yields the constraints
\begin{align}
iia) \ & \lambda_1 > 0 \, , \\
iib) \ & \lambda_2 > 0 \, , \\
iic) \ & \lambda_{12} + \zeta_4 \lambda_4 + 2 \sqrt{\lambda_1 \lambda_2} > 0 \, . 
\end{align}
In particular, since $\zeta_4 \in [0,1]$, $iic)$ leads (by considering $\lambda_4$ either positive or negative) to
\begin{align}
iic1) \ & \lambda_{12}  + 2 \sqrt{\lambda_1 \lambda_2} > 0 \, , \\
iic2) \ & \lambda_{12} + \lambda_4 + 2 \sqrt{\lambda_1 \lambda_2} > 0 \, . 
\end{align}
We are hence left with discussing $iii)$. 
{Although we were not able to solve it analytically}, 
it is still useful to consider some specific field directions. 
For instance, for $\phi = 0,\pi/2,\pi/4$ we get, respectively: 
\begin{align}
iiia) \    \lambda_{13}  &+ 2 \sqrt{\lambda_1 \lambda_3} > 0 \, , \\
iiib) \    \lambda_{23}  &+ 2 \sqrt{\lambda_2 \lambda_3} > 0 \, , \\
iiic) \    \lambda_{13} &+ \lambda_{23} + 2 \zeta_5 \lambda_5  \nn\\
&+ 
2 \sqrt{\lambda_3 \left( \lambda_1 + \lambda_2  + \lambda_{12} + \zeta_4 \lambda_4 \right) } > 0 \, . 
\end{align}
Notice that $\phi = \pi/4$ maximizes the contribution to the invariant $\zeta_5$. 
However this choice does not need to yield the most restrictive condition on $\lambda_5$, since other couplings 
might be magnified in other directions. 
By taking into account the uncorrelated {ranges} of $\zeta_4 \in [0,1]$ and $\zeta_5 \in [-1,1]$ (thus leading to a sufficient condition), 
we can further expand $iiic)$ into
\begin{align}
iiic1) \  \left| \lambda_5 \right| < &\ \frac{1}{2} (\lambda_{13} + \lambda_{23}) 
+ \sqrt{\lambda_3 \left( \lambda_1 + \lambda_2  + \lambda_{12}  \right) } \, , \\
iiic2) \  \left| \lambda_5 \right| <  &\ \frac{1}{2} (\lambda_{13} + \lambda_{23}) \nn\\
&+ \sqrt{\lambda_3 \left( \lambda_1 + \lambda_2  + \lambda_{12} + \lambda_4 \right) }  \, .
\end{align}

\subsection{Type-II seesaw limit: $(\theta, \phi) = (\frac{\pi}{2}, 0)$ 
}
\label{vacstaTII}

This limit corresponds to an effective Type-II seesaw model where only the field directions $\Delta$ and $H_u$ are considered in the {potential, which} reads
\begin{align}
V_0 = r^4 \cos^4\chi &\left[ ( \lambda_{\Delta 4}\right. + \zeta_9 \lambda_9 ) \nn\\
&+  \left.\left( \lambda_{\Delta 1} + \zeta_7 \lambda_7 \right) \tan^2\chi + \lambda_1 \tan^4\chi \right] \, .
\end{align}
The positivity constraints are hence
\begin{align} 
i) \ & \lambda_1 > 0 \, , \\
ii) \ & \lambda_{\Delta 4} + \zeta_9 \lambda_9 > 0 \, , \\
iii) \ & \lambda_{\Delta 1} + \zeta_7 \lambda_7 + 2 \sqrt{\lambda_1 \left( \lambda_{\Delta 4} + \zeta_9 \lambda_9 \right)} > 0 \, .
\end{align}
Given the (uncorrelated) domains of $\zeta_{7} \in [0,1]$ and $\zeta_{9} \in [\tfrac{1}{2},1]$ we find the (sufficient) conditions \cite{Arhrib:2011uy} 
\begin{align} 
iia) \ & \lambda_{\Delta 4} + \tfrac{1}{2} \lambda_9 > 0 \, , \\
iib) \ & \lambda_{\Delta 4} + \lambda_9 > 0 \, , 
\end{align}
and 
\begin{align} 
iiia) \ & \lambda_{\Delta 1} + \lambda_7 + 2 \sqrt{\lambda_1 \left( \lambda_{\Delta 4} + \tfrac{1}{2} \lambda_9 \right)} > 0 \, , \\
iiib) \ & \lambda_{\Delta 1} + 2 \sqrt{\lambda_1 \left( \lambda_{\Delta 4} + \tfrac{1}{2} \lambda_9 \right)} > 0 \, , \\
iiic) \ & \lambda_{\Delta 1} + \lambda_7 + 2 \sqrt{\lambda_1 \left( \lambda_{\Delta 4} + \lambda_9 \right)} > 0 \, , \\
iiid) \ & \lambda_{\Delta 1} + 2 \sqrt{\lambda_1 \left( \lambda_{\Delta 4} + \lambda_9 \right)} > 0 \, .
\end{align}

Analogous results are obtained by projecting along the $\Delta$--$H_d$ direction $(\theta, \phi) = (\frac{\pi}{2}, \frac{\pi}{2})$, 
which entails the replacements $\lambda_1 \rightarrow \lambda_2$, $\lambda_7 \rightarrow \lambda_8$ and  $\lambda_{\Delta 1} \rightarrow \lambda_{\Delta 2}$ 
in conditions $iiia$--$d)$.

\subsection{2HDM Type-II seesaw limit: $(\theta = \frac{\pi}{2})$}

In {this limit the singlet is decoupled and} we get 
\begin{widetext}
\begin{align}
V_0 &= r^4 \cos^4\chi 
\left\{ 
\left( \lambda_{\Delta 4} + \zeta_9 \lambda_9 \right) + \left[ \left( \lambda_{\Delta 1} + \zeta_7 \lambda_7 \right) \cos^2\phi 
+ \left( \lambda_{\Delta 2} + \zeta_8 \lambda_8 \right) \sin^2\phi
\right] \tan^2\chi \right. \nonumber \\
& \left.
+ \left[ \lambda_1 \cos^4\phi + \left( \lambda_{12} + \zeta_4 \lambda_4 \right) \cos^2\phi \sin^2\phi 
+ \lambda_2 \sin^4\phi 
\right] \tan^4\chi 
\right\} \, ,
\end{align}
which yields the positivity conditions 
\begin{align} 
i) \ & \lambda_{\Delta 4} + \zeta_9 \lambda_9 > 0 \, , \\
ii) \ & \lambda_1 \cos^4\phi + \left( \lambda_{12} + \zeta_4 \lambda_4 \right) \cos^2\phi \sin^2\phi + \lambda_2 \sin^4\phi > 0 \, , \\
iii) \ & \left( \lambda_{\Delta 1} + \zeta_7 \lambda_7 \right) \cos^2\phi 
+ \left( \lambda_{\Delta 2} + \zeta_8 \lambda_8 \right) \sin^2\phi  \nonumber \\
&+ 2 \sqrt{\left( \lambda_{\Delta 4} + \zeta_9 \lambda_9 \right) \left( \lambda_1 \cos^4\phi + \left( \lambda_{12} + \zeta_4 \lambda_4 \right) \cos^2\phi \sin^2\phi + \lambda_2 \sin^4\phi  \right) } > 0 \, .
\end{align}
\end{widetext}
{Note that condition $i)$} is equivalent to  
\begin{align} 
ia) \ & \lambda_{\Delta 4} + \tfrac{1}{2} \lambda_9 > 0 \, , \\
ib) \ & \lambda_{\Delta 4} + \lambda_9 > 0 \, , 
\end{align}
while $ii)$ {yields}
\begin{align} 
iia) \ & \lambda_1 > 0 \, , \\
iib) \ & \lambda_2 > 0 \, , \\
iic) \ & \lambda_{12} + \zeta_4 \lambda_4 + 2 \sqrt{\lambda_1 \lambda_2} > 0 \, .
\end{align}
The latter can be further expanded into
\begin{align} 
iic1) \ & \lambda_{12} + 2 \sqrt{\lambda_1 \lambda_2} > 0 \, , \\
iic2) \ & \lambda_{12} + \lambda_4 + 2 \sqrt{\lambda_1 \lambda_2} > 0 \, .
\end{align}
Condition $iii)$ for $\phi = 0$ and  $\frac{\pi}{2}$ has been already considered in the previous section.

\subsection{Single Higgs doublet limit: $(\phi=0)$}

In this case {$H_d$ is decoupled from the scalar potential, which} reads
\begin{align}
V_0 &= r^4 \cos^4\chi 
\left\{ 
\left( \lambda_{\Delta 4} + \zeta_9 \lambda_9 \right) \right.  \\
&+ \left[ \lambda_{\Delta 3} \cos^2\theta 
+ \left( \lambda_{\Delta 1} + \zeta_7 \lambda_7 \right) \sin^2\theta
\right] \tan^2\chi  \nonumber \\
& \left.
+ \left[ \lambda_3 \cos^4\theta + \lambda_{13} \cos^2\theta \sin^2\theta
+ \lambda_1 \sin^4\theta 
\right] \tan^4\chi 
\right\} \, , \nonumber
\end{align}
thus yielding the constraints
\begin{widetext}
\begin{align} 
i) \ & \lambda_{\Delta 4} + \zeta_9 \lambda_9 > 0 \, , \\
ii) \ & \lambda_3 \cos^4\theta + \lambda_{13} \cos^2\theta \sin^2\theta
+ \lambda_1 \sin^4\theta > 0 \, , \\
iii) \ & \lambda_{\Delta 3} \cos^2\theta 
+ \left( \lambda_{\Delta 1} + \zeta_7 \lambda_7 \right) \sin^2\theta 
+ 2 \sqrt{\left( \lambda_{\Delta 4} + \zeta_9 \lambda_9 \right) \left( \lambda_3 \cos^4\theta + \lambda_{13} \cos^2\theta \sin^2\theta
+ \lambda_1 \sin^4\theta  \right) } > 0 \, .
\end{align}
\end{widetext}

After further expanding $i)$ and $ii)$, we get

\begin{align} 
ia) \ & \lambda_{\Delta 4} + \tfrac{1}{2} \lambda_9 > 0 \, , \\
ib) \ & \lambda_{\Delta 4} + \lambda_9 > 0 \, , 
\end{align}

\begin{align} 
iia) \ & \lambda_3 > 0 \, , \\
iib) \ & \lambda_1 > 0 \, , \\
iic) \ & \lambda_{13} + 2 \sqrt{\lambda_1 \lambda_3} > 0 \, .  
\end{align}

The case $\phi = \frac{\pi}{2}$ ($H_u$ decoupled) is obtained by replacing $\lambda_1 \rightarrow \lambda_2$ and $\lambda_{13} \rightarrow \lambda_{23}$  
in the inequalities above.


\bibliographystyle{utphysmod}
\bibliography{sinupq}

\end{document}